\documentclass[journal,twoside,web]{IEEEtran}
\usepackage{silence}
\usepackage[font=scriptsize]{caption}
\usepackage{booktabs, makecell, multirow, tabularx}
\usepackage{cite}
\usepackage{amsmath,amssymb,amsfonts}
\usepackage{algorithmic}
\usepackage{graphicx}
\usepackage{textcomp}
\usepackage{hyperref}
\usepackage[utf8]{inputenc}
\usepackage[T1]{fontenc}
\usepackage{color,soul}
\usepackage{times}
\usepackage[nointegrals]{wasysym} 
\usepackage{epsfig}
\usepackage{multicol,multirow}
\usepackage{booktabs}
\usepackage{bigstrut}
\usepackage{floatrow}
\usepackage{rotating}
\usepackage{adjustbox}
\usepackage{mathtools}
\usepackage{longtable}
\usepackage{array}
\usepackage{pifont}

\def\BibTeX{{\rm B\kern-.05em{\sc i\kern-.025em b}\kern-.08em

    T\kern-.1667em\lower.7ex\hbox{E}\kern-.125emX}}

\begin{document}
\title{TBConvL-Net: A Hybrid Deep Learning Architecture for Robust Medical Image Segmentation}




\author{Shahzaib~Iqbal,
        Tariq~M.~Khan,
        Syed~S.~Naqvi,
        Asim~Naveed,
        and Erik Meijering,
\thanks{Shahzaib Iqbal and Syed S. Naqi are with the Department of Electrical and Computer Engineering, COMSATS University Islamabad (CUI), Islamabad, Pakistan.}
\thanks{Tariq M. Khan and Erik Meijering are with the School of Computer Science and Engineering, University of New South Wales, Sydney, NSW, Australia.}
\thanks{Asim Naveed is with the Department of Computer Science and Engineering, University of Engineering and Technology (UET) Lahore, Narowal Campus, Pakistan}}
\maketitle

\begin{abstract}
Deep learning has shown great potential for automated medical image segmentation to improve the precision and speed of disease diagnostics. However, the task presents significant difficulties due to variations in the scale, shape, texture, and contrast of the pathologies. Traditional convolutional neural network (CNN) models have certain limitations when it comes to effectively modelling multiscale context information and facilitating information interaction between skip connections across levels. To overcome these limitations, a novel deep learning architecture is introduced for medical image segmentation, taking advantage of CNNs and vision transformers. Our proposed model, named TBConvL-Net, involves a hybrid network that combines the local features of a CNN encoder-decoder architecture with long-range and temporal dependencies using biconvolutional long-short-term memory (LSTM) networks and vision transformers (ViT). This enables the model to capture contextual channel relationships in the data and account for the uncertainty of segmentation over time. Additionally, we introduce a novel composite loss function that considers both the segmentation robustness and the boundary agreement of the predicted output with the gold standard. Our proposed model shows consistent improvement over the state of the art on ten publicly available datasets of seven different medical imaging modalities.

\end{abstract}

\begin{IEEEkeywords}
Medical Image Segmentation, CNN, LSTM, Vision Transformers
\end{IEEEkeywords}

\section{Introduction}
\label{INTRO}
The accurate segmentation of lesions and other pathologies in medical images poses a significant challenge, but remains a crucial task in the field of medical image analysis \cite{khan2019boosting,khan2020exploiting,iqbal2022g}. Relying solely on expert opinions for diagnosis can be time-consuming and subject to bias from clinical experience \cite{iqbal2022recent, khan2021residual, khan2021rc, khan2022t,javed2024region,iqbal2024tesl}. Hence, automated medical image segmentation (MIS) can be greatly valuable for medical professionals and can offer substantial advantages for disease diagnosis and treatment planning \cite{arsalan2022prompt,khan2022neural,khan2022mkis,naveed2024ra,khan2024lmbf}. In the field of computer vision, convolutional neural networks (CNNs) have gained prominence as the prevailing segmentation method\cite{naqvi2023glan,khan2023retinal,manan2023semantic,khan2024esdmr}. This is evident from the extensive use of CNN architectures, such as deep residual networks \cite{he2016deep}, DenseNet \cite{huang2017densely}, and EfficientNet \cite{tan2019efficientnet}. Similarly, in medical image analysis, CNNs such as Ce-Net \cite{gu2019net}, FES-Net \cite{khan2023feature}, M-Net \cite{fu2018joint}, MLR-Net \cite{iqbal2023mlr}, LDMRes-Net \cite{iqbal2023ldmres}, LMBiS-Net \cite{abbasi2023lmbis}, U-Net \cite{ronneberger2015u} and U-Net ++ \cite{zhou2019unet++} have attracted significant attention and application. Most segmentation methods commonly use U-Net \cite{ronneberger2015u} or its variants \cite{zhou2019unet++, iqbal2022g,iqbal2024euis, khan2020semantically,khan2021residual}. However, the localised nature of convolutional operations in CNNs imposes constraints on their capacity to capture long-range dependencies, which can lead to less-than-optimal segmentation outcomes. This leads to two notable drawbacks. First, the utilisation of small convolutional kernels focusses mainly on local features, neglecting the importance of global features. Global features are crucial for reliably segmenting medical images with varying lesion shapes and sizes. Also, once they have been trained, the convolutional kernels cannot change based on the content of the input image. This makes the network less adaptable to different input features.
 
Self-attention-based transformers \cite{vaswani2017attention} have gained prominence in natural language processing, and their application to computer vision has attracted interest. Vision Transformers (ViT) \cite{dosovitskiy2020image} emerged as the pioneering approach that used transformer encoders for image classification. ViT did as well or better than CNN-based models, showing that self-attention mechanisms could be useful for computer vision. Transformers have also been used for other visual tasks, such as object detection \cite{Carion2020end} and semantic segmentation \cite{zheng2021rethinking}, with state-of-the-art (SOTA) performance showing the best results to date. In MIS, TransUNet \cite{chen2021transunet} was the pioneer model to incorporate a hybrid architecture consisting of CNN and transformers. Since then, transformer encoder-decoder models that are entirely based on transformers, such as Swin-UNet \cite{cao2023swin} and nnFormer \cite{zhou2021nnformer}, have been suggested to segment volumetric medical images. These approaches have shown strong performance because of their ability to capture interactions over long distances and dynamically encode features.

Although transformers have shown great success in modelling long-range dependencies and have been applied to MIS, they still have limitations. One of the drawbacks is that transformers tend to ignore crucial spatial and local feature information. Another drawback is that they require large datasets for training \cite{lee2021vision}, which limits their ability to model local visual cues \cite{xu2021vitae}. It should also be noted that transformers, despite their strengths, are limited in their ability to learn features using a token-wise attention mechanism on a single scale. Because of this limitation, transformers cannot easily record feature dependencies between channels at different scales, which can be a problem when working with pathologies that are different in size and shape. Consequently, there exists a potential for hybrid CNN-transformer architectures for MIS, as CNNs and transformers have complementary strengths. CNNs are data efficient and suitable for preserving local spatial information. Transformers, on the other hand, can model long-range dependencies and perform dynamic attention, making them useful for segmenting large-scale lesions. Previous work \cite{wu2021cvt, zhang2021rest} has attempted to combine these two types of model for feature encoding, but with high computational complexity and reliance on large-scale datasets such as ImageNet. Moreover, current hybrid techniques employ only a single token granularity in each attention layer, disregarding the channel relationships of transformers and their importance in feature extraction.

\begin{figure*}[h]
  \centering
  \vspace{0.5cm}
  \includegraphics[scale=0.27]{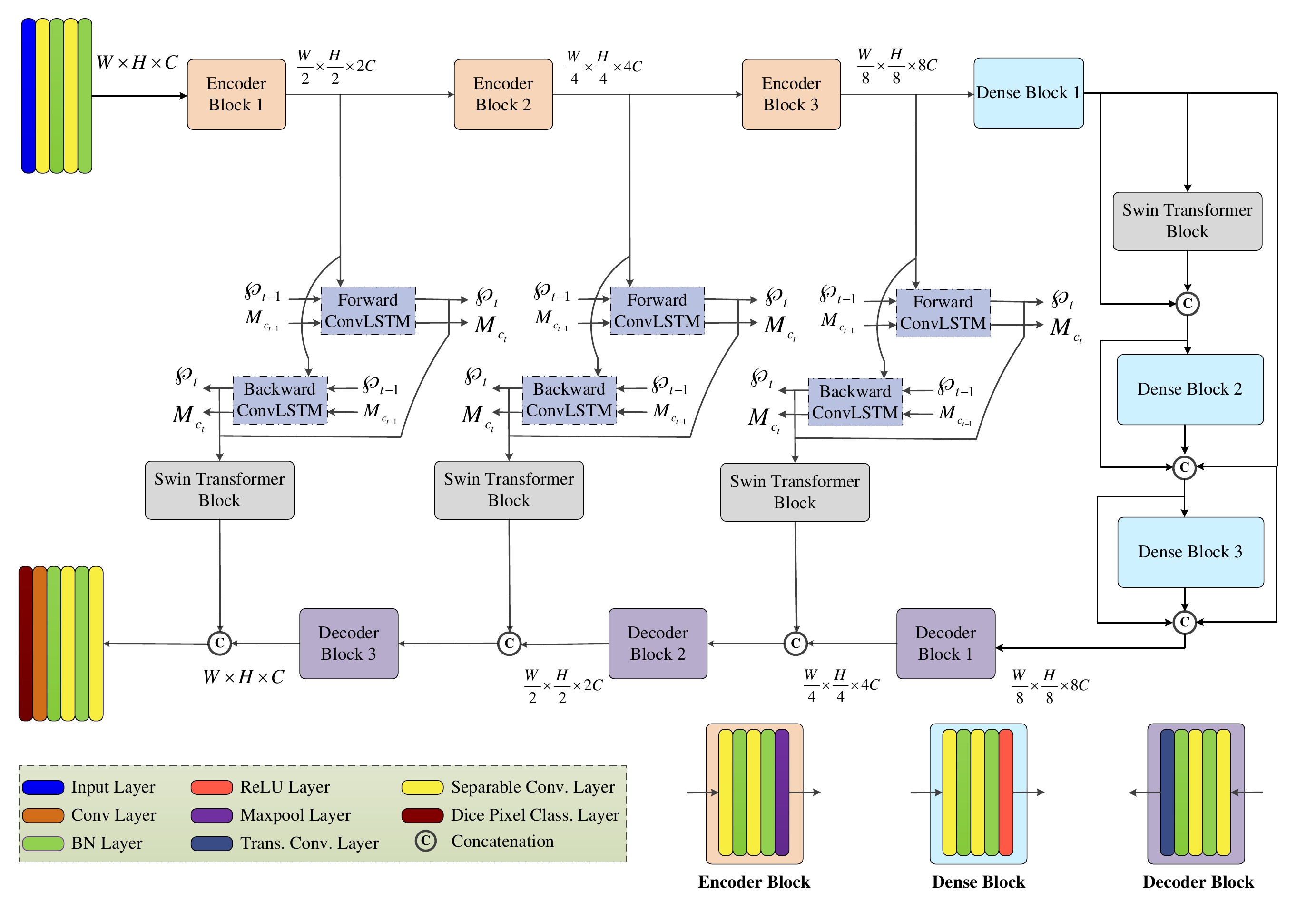}\\
 \caption{Block diagram of the TBConvL-Net architecture, showing its key components: encoder, decoder, and skip connections with BConvLSTM and Transformer layers.}
 \label{Fig: model}
\end{figure*}

\begin{figure*}[h]
  \centering
  \includegraphics[scale=0.35]{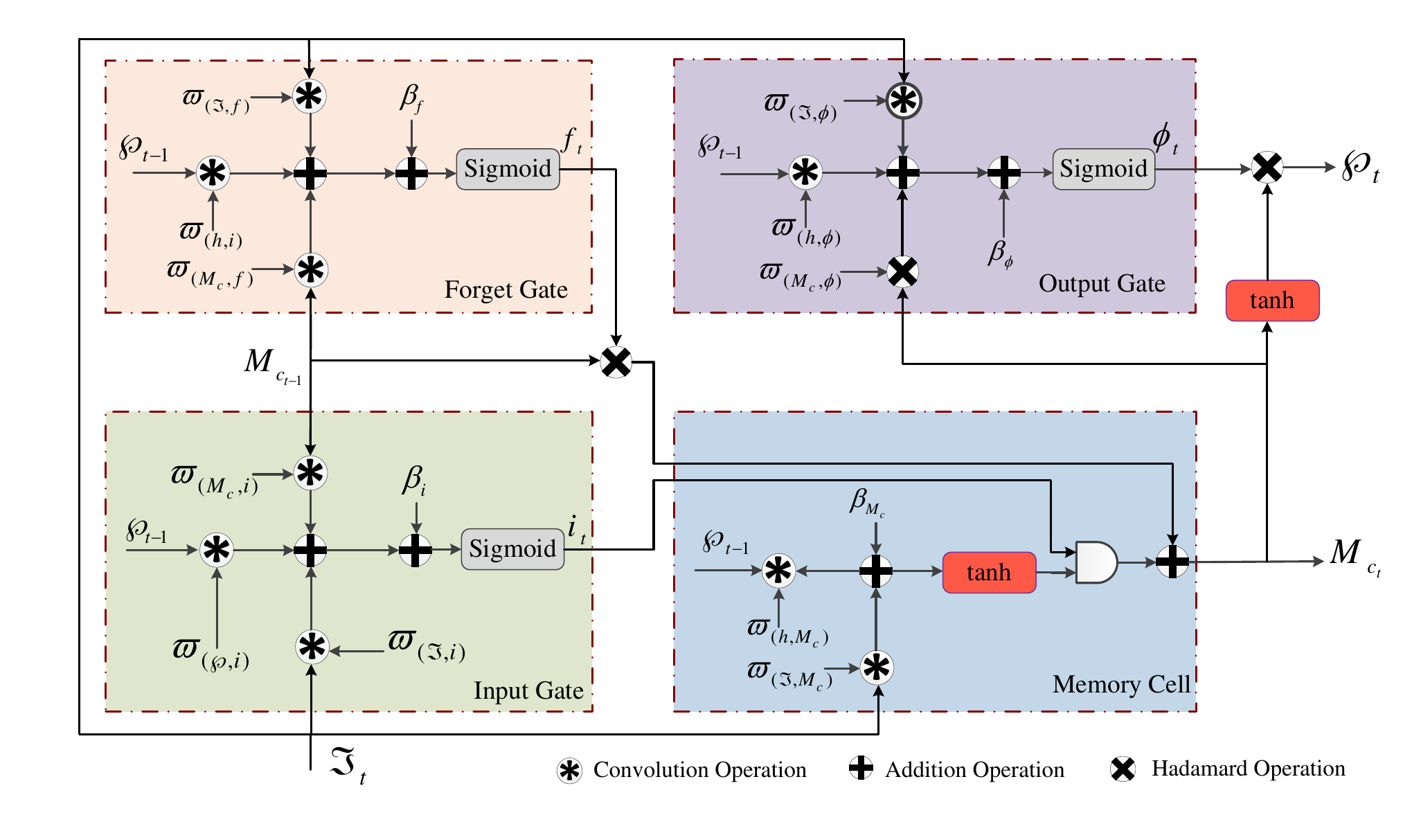}\\
 \caption{Design of the ConvLSTM block, a solution to the spatial correlation shortcomings of traditional LSTM models, achieved by the incorporation of convolutional operations in the input-to-state and state-to-state transitions. The architecture includes a memory cell ($M_c$), an output gate ($\phi$), an input gate ($i$) and a forget gate ($f$), with these gates serving as control mechanisms to access, update and erase the content of the memory cells. For both the hidden and the input states in the block, 2D convolution masks are used, with Hadamard and convolutional operations symbolised by $\otimes$ and $\circledast$, respectively. The input and hidden state tensors are indicated by $\Im_t$ and $\wp_t$, respectively, while the biases associated with the memory cell, the output gate, the input gate and the forget gate are denoted as $\beta_{M_c}$, $\beta_{\phi}$, $\beta_i$, and $\beta_f$, respectively.}
 \label{Fig:CONVLSTM}
\end{figure*}

To overcome these limitations, it is necessary to explore effective ways to integrate the strengths of CNNs and transformers for MIS while maintaining computational efficiency and avoiding their respective drawbacks. Here, we introduce TBConvL-Net, a novel architecture that combines the strengths of CNNs and transformers with bidirectional long-short-term memory (LSTM) models specifically designed for MIS. The encoder part consists of several hierarchical separable convolutional layers of CNNs, responsible for capturing the spatial information in the input image. In the decoder section of the architecture, multiple separable convolutional layers and upsampling layers are used to facilitate the reconstruction process. Bridging the semantic and resolution gap between encoder and decoder features is crucial to capture the multi-scale global context in MIS. Specifically, encoder features have higher resolution, enabling them to capture more fine-grained information, while decoder features possess higher semantic information and contextual understanding. Hence, the objective is to learn the transfer of multi-scale contextual information while preserving the integrity of the semantic information. We aim to achieve this without adversely impacting the richness and accuracy of contextual understanding embedded within the decoder features by using a combination of bidirectional ConvLSTM (BConvLSTM) and transformers within the skip connections of the proposed TBConvL-Net. This enables robust feature fusion in the encoder-decoder architecture, marking the first instance of such an application. The key idea is to leverage the power of vision transformers for contextual processing in the spatial domain while considering the temporal interactions between features for semantically aware feature fusion. However, the challenge in employing traditional transformers in dense prediction tasks is the quadratic computational complexity of self-attention. To reduce complexity and improve modelling of long-range dependencies and robustness to image variations, we introduce a lightweight Swin Transformer Block (STB) in skip connections for semantically aware feature fusion \cite{liu2021swin}. The shifted windowing-based self-attention layers model long-range dependencies and dynamic attention across the image, allowing the model to capture features at different scales and encode channel relationships. BConvLSTM complements the transformer in learning the forward and backward temporal dependencies and patterns between the encoder and the decoder features. 

Compared to existing methods, the proposed TBConvL-Net features the following innovations, facilitating the MIS task. First, the design of a hybrid network that takes into account the local features of a CNN encoder-decoder architecture, as well as temporal and long-range dependencies through BConvLSTM and Swin Transformer, allows one to account for segmentation uncertainties over time and captures contextual channel relationships in the data. Second, the composite loss function considers both the robustness of the segmentation and the boundary agreement of the predicted output with the gold standard. Third, the use of depth-wise separable convolutions instead of traditional convolutions minimises computational burden and improves feature learning by exploiting filter redundancy. Using the optimal number of filters prevents filter overlap and promotes convergence to globally optimal minima. The proposed method is evaluated for seven medical image segmentation applications using ten public datasets. Tasks include thyroid nodule segmentation, breast cancer lesion segmentation, optic disc segmentation, chest radiograph segmentation, nuclei cell segmentation, fluorescent neuronal cell segmentation, and skin lesion segmentation. Our experimental results show that our method consistently outperforms current SOTA methods while also requiring fewer computational resources. Therefore, the method offers great benefits for segmenting medical images with limited resources.

\begin{figure*}[t]
  \centering
  \vspace{0.5cm}
  \includegraphics[scale=0.90]{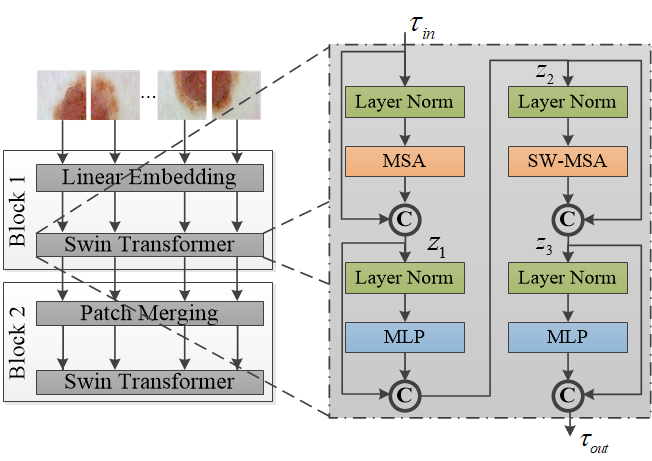}\\
 \caption{Lightweight swin transformer architecture. The input RGB images are divided into non-overlapping patches, transformed into tokens, and projected into an arbitrary dimension ($d$). Transformer blocks with modified self-attention computations process these tokens, creating a hierarchical representation. The lightweight version replaces the conventional multihead self-attention (MSA) module with a shifted window-based MSA module to reduce computational complexity while preserving core functionality. Efficiency is further improved by computing self-attention within local windows, scaling linearly with a fixed size of $N$.}
 \label{Fig: tranformer}
\end{figure*}

\section{Related Work}
\label{sec: Related Work}

CNNs, a form of deep learning model, have seen substantial use and recognition in the field of MIS. This is due to their outstanding ability to extract image features efficiently. Among the notable architectures, U-Net \cite{ronneberger2015u} has emerged as a pioneering model, exhibiting competitive performance in various MIS tasks. Based on U-Net, several variants have been proposed, including UNet++ \cite{zhou2019unet++}, nnUNet \cite{isensee2021nnu}, UNet3+ \cite{huang2020unet}, Dense-UNet \cite{li2018h}, and Attention U-Net \cite{oktay2018attention}. These variants of U-Net and customised approaches demonstrate the adaptability and effectiveness of CNNs in addressing various challenges in MIS tasks, catering to specific anatomical structures, diseases, or imaging modalities.

Transformer-based methods have also shown remarkable performance in various vision tasks \cite{zheng2021rethinking, zhang2021rest, liu2021swin}. The ViT architecture \cite{vaswani2017attention} revolutionised the application of transformers in image classification, showcasing their efficacy in the capture of global contextual information. Subsequent advances, such as the DeiT model \cite{touvron2021training}, have introduced efficient training strategies to improve ViT performance. A notable development is the Swin Transformer \cite{liu2021swin}, which uses self-attention with local windows, allowing for more computationally efficient processing while still achieving satisfactory results. To combine the strengths of CNNs and transformers, some approaches have incorporated the design principles of the former into the latter. For example, CoatNet \cite{dai2021coatnet} and Bottleneck Transformers \cite{srinivas2021bottleneck} introduced CNN-inspired design elements into transformers, resulting in improved performance and resource efficiency. These advances in transformer-based methods and their demonstrated potential in vision tasks provide avenues to explore their effectiveness for MIS.

In MIS, many approaches have been developed to address 2D and 3D tasks. These approaches aim to address challenges specific to medical image data \cite{iqbal2023robust} and encompass various techniques and methodologies. These include methods such as nnFormer \cite{zhou2021nnformer}, TransUNet \cite{chen2021transunet}, and others \cite{he2023h2former, yan2022after}. TransUNet \cite{chen2021transunet} was the first to merge the strengths of CNN and transformer architectures for MIS. This innovative model leverages CNN's capacity to extract local features while benefiting from the global contextual feature recognition provided by transformers. To mitigate the data-intensive requirement associated with transformers, UTNet \cite{gao2021utnet} was introduced. This method incorporates a self-attention mechanism into a CNN framework, which results in enhanced performance in MIS tasks. However, TransUNet and UTNet are more prone to overfitting due to their complex architectures and redundant feature learning and are more computationally demanding in the training phase. Based on the ideas of the Swin Transformer \cite{liu2021swin}, the Swin-UNet model \cite{cao2023swin} was introduced. However, it does not pay significant attention to local spatial information, which is a critical factor in the segmentation process.

\begin{table*}[htbp]
  \centering
  \caption{Details of the medical image datasets used for evaluation.}
  \resizebox{1.0\textwidth}{!}{%
  
    \begin{tabular}{llccccccccl}
    \toprule
    \multirow{2}[4]{*}{\textbf{Modality}} & \multirow{2}[4]{*}{\textbf{Dataset}} & \multicolumn{4}{c}{\textbf{Image Count}} & \multirow{2}[4]{*}{\textbf{Image Resolution Range}} & \multirow{2}[4]{*}{\textbf{Format}} & \multirow{2}[4]{*}{\textbf{Resized}} & \multirow{2}[4]{*}{\textbf{Data Split}} & \multirow{2}[4]{*}{\textbf{Task}} \\
\cmidrule{3-6}          &       & \textbf{Training} & \textbf{Validation} & \textbf{Testing} & \textbf{Total} &       &       &       &       &  \\
    \midrule
    \multirow{3}[2]{*}{Optical Imaging} & ISIC 2016 \cite{gutman2016skin} & 900   & -     & 379   & 1279  & $679\times 453 - 6748\times 4499$ & JPEG & \multirow{3}[2]{*}{$256\times 256$} & \multirow{3}[2]{*}{-} & \multirow{3}[2]{*}{Skin Lesions Segmentation} \\
          & ISIC 2017 \cite{codella2018skin}  & 2000  & 150   & 600   & 2750  & $679\times 453 - 6748\times 4499$ & JPEG &       &       &  \\
          & ISIC 2018 \cite{codella2019skin} & 2594  & -     & 1000  & 3594  & $679\times 453 - 6748\times 4499$ & JPEG &       &       &  \\
          
    \multirow{2}[2]{*}{Ultrasound Imaging} & DDTI \cite{DDTI} & -     & -     & -     & 637   & $245\times 360 - 560\times 360$  & PNG  & \multicolumn{1}{l}{$256\times 256$} & \multicolumn{1}{l}{80\%:10\%:10\%} & Thyroid Nodule Segmentation \\
          & BUSI \cite{BUSIdataset} & -     & -     & -     & 780   & $319\times 473 - 583\times 1010$ & PNG  & \multicolumn{1}{l}{$256\times 256$} & \multicolumn{1}{l}{80\%:10\%:10\%} & Breast Ultrasound Segmentation (Age 25--75) \\
 
    WSI Imaging & MoNuSeg \cite{MoNuSeg_Dataset} & 30    & -     & 14    & 44    & $1000\times 1000$ & PNG  & $512\times 512$ & -     & Nuclei Segmentation \\
 
    X-Ray Imaging & MC \cite{MC_Dataset}    & 100   & -     & 38    & 138   & $4892\times 4020 - 4020\times 4892$ & TIF  & $512\times 512$ & -     & Chest X-Rays Segmentation \\
 
    Fundus Imaging & IDRiD \cite{IDRiD_Dataset}& 54    & -     & 27    & 81    & $4288\times 2848$ & JPEG & $512\times 512$ & -     & Optic Disc Segmentation \\
 
    Microscopic Imaging & Fluorescent Neuronal Cells \cite{Fluocell_Dataset} & 283   & -     & 70    & 353   & $1600\times 1200$ & PNG  & $512\times 512$ & -     & Fluorescent Microscopic Cells Segmentation \\
    MRI Imaging & The Cancer Imaging Archive (TCIA) \cite{BrainMRI_Dataset} & 1084  & -     & 285   & 1369  & $256\times 256$ & TIF  & $256\times 256$ & -     & Brain Tumour Segmentation \\

    \bottomrule
    \end{tabular}%
}
  \label{tab:Datasets}%
\end{table*}%

\section{Proposed Method}
\label{sec: Proposed Method}

TBConvL-Net consists of multiple key components (Fig.~\ref{Fig: model}). Here, we describe the encoder-decoder architecture, the BConvLSTM and transformer block, and the loss function of the proposed network.

\subsection{Encoder-Decoder Architecture}
\label{sec:overview}

The encoder component of TBConvL-Net consists of four stages, with each stage comprising two separable convolutional layers using $3\times3$ filters. This is followed by a $2\times2$ max pooling layer and the application of a Rectified Linear Unit (ReLU) activation function. With each subsequent stage, the number of filters doubles compared to the previous stage. By progressively increasing the layer dimensions, the TBConvL-Net encoder path gradually extracts visual features, culminating in the final layer generating high-level semantic information based on high-dimensional image representations.

Unlike legacy feature learning in CNNs, where independent feature learning is promoted in different layers, densely connected convolutions are proposed \cite{huang2017densely}. The concept of ``collective knowledge'' is used to improve network performance by reusing feature maps throughout the network. In this approach, the feature maps generated from earlier convolutional layers are integrated with those from the existing layer. This combined output is then fed into the subsequent convolutional layer. Densely connected convolutions have notable benefits over traditional convolutions \cite{huang2017densely}. First, they help the network learn a broad range of feature maps rather than redundant features. Additionally, feature reuse and information sharing throughout the network enhance the network's ability to represent complex features. Finally, as densely connected convolutions can benefit from every feature that has been formed before them, the network is able to avoid the danger of gradients bursting or vanishing.

\begin{table*}[h]
  \centering
  \caption{Results of the ablation study of the different locations of the transformer in the baseline network on the ISIC 2017 dataset. $^{*}$SC is a depth-wise separable convolution.}
      \adjustbox {max width=\textwidth}{%
    \begin{tabular}{llccccc}
    \toprule
    \multirow{2}[4]{*}{\textbf{Method}} & \multirow{2}[4]{*}{\textbf{Transformer Location in the Network}} & \multicolumn{5}{c}{\textbf{Performance (\%)}} \\
\cmidrule{3-7}          &       & \textbf{$J$} & \textbf{$D$ } & \textbf{$A_{cc}$} & \textbf{$S_{n}$} & \textbf{$S_{p}$} \\
    \midrule
    Baseline Model (BM) & Not applicable & 79.20 & 78.11 & 91.63 & 76.46 & \bf 97.09 \\
    BM with SC$^{*}$ (SC-BM) & Not applicable & 80.14 & 87.60 & 94.33 & 88.87 & 94.60 \\
    SC-BM + Swin Transformer & Between dense layer of the network & 78.61 & 86.45 & 93.78 & 87.07 & 95.11 \\
    SC-BM + Swin Transformer & After every pooling layer in the decoder & 81.53 & 88.88 & 94.73 & 89.28 & 94.83 \\
    SC-BM + Swin Transformer & Between the skip connections of the network & 81.70 & 88.79 & 94.47 & 90.20 & 94.01 \\
    SC-BM + Swin Transformer & Between the skip connections and dense layers of the network & \bf 82.78 & \bf 89.66 & \bf 95.07 & \bf 90.21 & 95.18 \\
    \bottomrule
    \end{tabular}%
}
  \label{tab:ablation 1}%
\end{table*}%
\begin{figure*}[h]
	\centering
    \includegraphics[width=\textwidth]{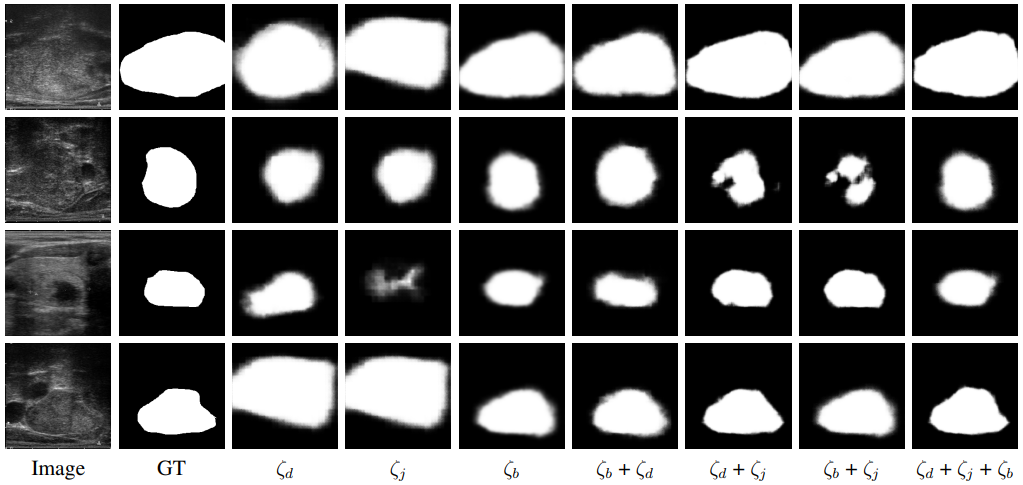} 
	\caption{Visual results of the proposed TBConvL-Net using the different loss functions for thyroid nodule segmentation in the DDTI dataset.}
	\label{Fig: DDTI Loss Ablation}%
\end{figure*}%

Let $l^{*\times *}$ denote the depth-wise separable convolution ($f_{s}^{*\times *}$) operation of any given kernel size ($*\times *$) followed by the batch normalisation ($\beta_{N}$) operation on any given input ($I$):
\begin{equation}
    l^{*\times *}=\beta_{N}({f_{s}^{*\times *}(I)}).
\label{Eq:1}
\end{equation}
Furthermore, let $B_{i}^\text{enc}$ be the output of the $i^\text{th}$ encoder block, where $i=1, 2, 3$, computed by applying two consecutive $l^{(3\times 3)}$ operations followed by the $i^\text{th}$ ($2\times 2$) max-pooling operation ($M_{P_{i}}$) on the encoded features ($\chi_\text{in}$):
\begin{equation}
    B_{i}^\text{enc}=M_{P_{i}}(l^{3\times 3}(l^{3\times 3}(\chi_\text{in}))).
\label{Eq:2}
\end{equation}
In TBConvL-Net, three encoding blocks are used with progressively smaller spatial input dimensions, namely $W\times H\times C$ for $B_{1}^\text{enc}$, $\frac{1}{2}W\times \frac{1}{2}H\times 2C$ for $B_{2}^\text{enc}$, and $\frac{1}{4}W\times \frac{1}{4}H\times 4C$ for $B_{3}^\text{enc}$. After the encoding blocks, three densely connected depth-wise separable convolution blocks $B_{i}^\text{den}$ are used with spatial input dimensions $\frac{1}{8}W\times \frac{1}{8}H\times 8C$. The output of the first dense block $B_{1}^\text{den}$ is calculated by applying the two consecutive operations $l^{(3\times 3)}$ followed by the activation function on the last encoding block $B_{3}^\text{enc}$:
\begin{equation}
    B_{1}^\text{den}=\Re(l^{3\times 3}(l^{3\times 3}(B_{3}^\text{enc}))),
\label{Eq:3}
\end{equation}
where $\Re$ is the activation function (ReLU). The output of the $2^\text{nd}$ dense block $B_{2}^\text{den}$ is calculated by applying STB ($S_\text{ViT}$) to $B_{1}^\text{den}$ and concatenating with it:
\begin{equation}
    B_{2}^\text{den}=S_\text{ViT}(B_{1}^\text{den})\copyright B_{1}^\text{den},
\label{Eq:4}
\end{equation}
where $\copyright$ denotes the concatenation operation. The output of the last densely connected depth-wise separable convolution block $B_{3}^\text{den}$ is computed by applying the two consecutive $l^{(3\times 3)}$ operations and concatenation of previous densely connected depth-wise separable convolution block $B_{1}^\text{den}$ and $B_{2}^\text{den}$:
\begin{equation}
    B_{3}^\text{den}=[\Re(l^{3\times 3}(l^{3\times 3}(B_{2}^\text{den})))]\copyright B_{1}^\text{den}\copyright B_{2}^\text{den}.
\label{Eq:5}
\end{equation}
In this process, two sequential operations are applied, each with a $3\times3$ filter, denoted as $l^{(3\times 3)}$. These operations are then concatenated with the outputs of the previous densely connected, depthwise separable convolution blocks, $B_{1}^\text{den}$ and $B_{2}^\text{den}$. This approach of merging the outputs of earlier blocks with the output of the current block enhances the ability of the network to learn more complex and high-level features. The decoder blocks are computed as:
\begin{equation}
 B_{i}^\text{dec}=T_{c_{i}}(l^{(3\times 3)}(l^{(3\times 3)}(B_{3}^\text{den})))\copyright S_\text{ViT}(\blacktriangle ^{\leftrightharpoons{}}_\text{lstm}(B_{i}^\text{enc})),
\label{Eq:6}
\end{equation}
where $T_{c_{i}}$ is the transposed convolution operation of the $i^\text{th}$ decoder block, $S_\text{ViT}$ is the STB, and $\blacktriangle ^{\leftrightharpoons{}}_\text{lstm}$ denotes the BConvLSTM. The final output of TBConvL-Net, $\chi_\text{out}$, is calculated by applying two consecutive $l^{(3\times 3)}$ operations followed by the sigmoid function $\varrho$ on the output of the last decoder block $B_{3}^\text{dec}$:
\begin{equation}
\chi_\text{out}=\varrho (l^{(3\times 3)}(l^{(3\times 3)}(B_{3}^\text{dec}))).
\label{Eq:7}
\end{equation}

\begin{table*}[h]
  \centering
  \caption{Results of the ablation study of the different loss functions in TBConvL-Net on thyroid nodule segmentation in the DDTI dataset.}
      \adjustbox {max width=\textwidth}{%
    \begin{tabular}{lcccccrccccc}
    \toprule
    \multirow{3}[5]{*}{\textbf{Loss Function}} & \multicolumn{11}{c}{\textbf{Performance Measures (\%)}} \\
\cmidrule{2-12}          & \multicolumn{5}{c}{\textbf{ISIC 2017}} &       & \multicolumn{5}{c}{\textbf{DDTI}} \\
\cmidrule{2-6}\cmidrule{8-12}          & \textbf{$J$} & \textbf{$D$ } & \textbf{$A_{cc}$} & \textbf{$S_{n}$} & \textbf{$S_{p}$} &       & \textbf{$J$} & \textbf{$D$ } & \textbf{$A_{cc}$} & \textbf{$S_{n}$} & \textbf{$S_{p}$} \\
\hline
    $\zeta_{d}$ & 78.72 & 87.26 & 94.27 & 86.44 & 95.00 &       & 76.88 & 86.17 & 97.14 & 84.05 & 97.07 \\
    $\zeta_{j}$ & 74.22 & 82.36 & 92.39 & 84.00 & 95.66 &       & 78.72 & 87.26 & 94.27 & 86.44 & 95.00 \\
    $\zeta_{b}$ & 63.06 & 73.21 & 90.24 & 75.91 & 96.25 &       & 74.22 & 82.36 & 92.39 & 84.00 & 95.66 \\
    $\zeta_{b}$ + $\zeta_{d}$ & 67.26 & 77.72 & 91.27 & 87.59 & 92.98 &       & 77.79 & 81.99 & 95.22 & 82.83 & 95.22 \\
    $\zeta_{d}$ + $\zeta_{j}$ & 82.83 & 89.71 & 95.07 & 90.08 & 95.41 &       & 81.22 & 82.88 & 94.85 & 82.91 & 96.99 \\
    $\zeta_{b}$ + $\zeta_{j}$ & 75.03 & 83.94 & 92.96 & 90.92 & 93.11 &       & 79.36 & 80.54 & 95.88 & 85.85 & 96.46 \\
    $\zeta_{d}$ + $\zeta_{j}$ + $\zeta_{b}$ & 83.91 & 90.57 & 95.64 & 92.68 & 96.80 &       & 86.06 & 89.90 & 95.45 & 90.26 & 97.71 \\
    \bottomrule
    \end{tabular}%
}
  \label{tab:ablation 2}%
\end{table*}%

\begin{table}[h]
  \centering
  \caption{Learnt weights transfer learning of the TBConvL-Net on different MIS datasets.}
   \adjustbox {max width=\textwidth}{%
    
    \begin{tabular}{ll}
    \toprule
    \textbf{Dataset} & \textbf{Transfer Learning} \\
    \midrule
    ISIC 2016 \cite{gutman2016skin} & Learnt weights of ISIC 2017 \\
    ISIC 2017 \cite{codella2018skin}& Learnt weights of ISIC 2018  \\
    ISIC 2018 \cite{codella2019skin} & Learnt weights of ISIC 2017  \\
    DDTI \cite{BUSIdataset} & Learnt weights of BUSI \\
    BUSI \cite{BUSIdataset} & Learnt weights of DDTI \\
    MoNuSeg \cite{MoNuSeg_Dataset}& Learnt weights of Fluorescent Neuronal Cells  \\
    MC \cite{MC_Dataset} & Without Transfer Learning \\
    IDRiD \cite{IDRiD_Dataset}& Without Transfer Learning \\
    Fluorescent Neuronal Cells \cite{Fluocell_Dataset}& Learnt weights of MoNuSeg \\
    TCIA \cite{BrainMRI_Dataset} & Learnt weights of MoNuSeg \\
    \bottomrule
    \end{tabular}%
    }
  \label{tab:trainig_strategy}%
\end{table}%

\begin{table}[h]
  \centering
  \caption{Performance enhancement achieved by TBConvL-Net by using the transfer learning strategy on different datasets of MIS.}
  \adjustbox {max width=0.9\textwidth}{%
    \begin{tabular}{lcccccc}
    \toprule
    \multirow{2}[4]{*}{\textbf{Dataset}} & \multirow{2}[4]{*}{\textbf{Transfer Learning}} & \multicolumn{5}{c}{\textbf{Performance Measures in (\%)}} \\
    \cmidrule{3-7} & & \textbf{$J$} & \textbf{$D$ } & \textbf{$A_{cc}$} & \textbf{$S_{n}$} & \textbf{$S_{p}$} \\
    \midrule
    \multirow{2}[2]{*}{ISIC 2016} & No & 86.10 & 91.76 & 96.32 & 93.57 & 94.88 \\
          & \textbf{Yes} & \textbf{89.47} & \textbf{95.45} & \textbf{97.05} & \textbf{94.02} & \textbf{97.68} \\
    \midrule
    \multirow{2}[2]{*}{ISIC 2017} & No & 81.78 & 89.66 & 95.07 & 90.21 & 95.18 \\
          & \textbf{Yes} & \textbf{84.80} & \textbf{90.89} & \textbf{96.07} & \textbf{91.19} & \textbf{97.61} \\
    \midrule
    \multirow{2}[2]{*}{ISIC 2018} & No & 87.31 & 92.54 & 96.04 & 91.94 & 97.61 \\
          & \textbf{Yes} & \textbf{91.65} & \textbf{95.47} & \textbf{97.60} & \textbf{95.29} & 98.55 \\
    \midrule
    \multirow{2}[2]{*}{DDTI} & No & 86.06 & 89.90 & 95.45 & 90.26 & 97.71 \\
          & \textbf{Yes} & \textbf{88.70} & \textbf{93.56} & \textbf{98.62} & \textbf{94.02} & \textbf{99.09} \\
    \midrule
    \multirow{2}[2]{*}{BUSI} & No & 85.95 & 91.42 & 96.92 & 92.82 & 95.24 \\
          & \textbf{Yes} & \textbf{91.97} & \textbf{95.72} & \textbf{99.50} & \textbf{95.85} & \textbf{99.69} \\
    \midrule
    \multirow{2}[2]{*}{MoNuSeg} & No & 70.59 & 81.34 & 94.22 & 85.38 & 95.24 \\
          & \textbf{Yes} & \textbf{76.07} & \textbf{85.16} & \textbf{93.62} & \textbf{88.04} & \textbf{95.53} \\
    \midrule
    \multirow{2}[2]{*}{MC} & No & 97.88  & \textbf{98.97}  & 99.50  & 98.40  & \textbf{99.05}\\
          & \textbf{Yes} & \textbf{97.90} & 98.86 & \textbf{99.50} & \textbf{97.69} & 99.04 \\
    \midrule
    \multirow{2}[2]{*}{IDRiD} & No & 95.65 &   96.73 & \textbf{99.94}  & \textbf{97.68}  & \textbf{99.97}\\
          & \textbf{Yes} & \textbf{95.67} & \textbf{96.73} & 99.93 & 97.62 & 99.95 \\
    \midrule
    \multirow{2}[2]{*}{Fluorescent Neuronal Cells} & No & 88.11 & 93.54 & 98.24 & 95.32 & 99.19 \\
          & \textbf{Yes} & \textbf{92.84} & \textbf{96.23} & \textbf{99.90} & \textbf{97.01} & \textbf{99.94} \\
    \midrule
    \multirow{2}[2]{*}{TCIA} & No & 88.71 & 94.55 & 97.15 & 94.22 & 97.86 \\
          & \textbf{Yes} & \textbf{92.93} & \textbf{95.47} & \textbf{99.34} & \textbf{95.63} & \textbf{99.79} \\
    \bottomrule
    \end{tabular}%
    }
  \label{tab:Ablation 3}%
\end{table}%

\subsection{Bidirectional ConvLSTM and Transformer Block}
\label{sec:Transformer}
By modelling long-range dependencies in both directions, bidirectional LSTMs can capture contextual information from past and future steps in the sequence. This can enhance the network's ability to learn complex patterns and relationships in the data. On the other hand, Swin Transformers use a hierarchical approach to process nonoverlapping local image patches, allowing them to learn features at various scales. This significantly enhances the ability of the network to model complex structures and relationships in the data. In the Swin Transformer architecture, the attention mechanism is employed both across patches and within them. This enables capturing global relationships among different parts of the input data. By considering both local and global dependencies, the Swin Transformer effectively learns the contextual information necessary for various tasks.

\begin{table*}[h]
  \centering
  \caption{Performance comparison of TBConvL-Net with various SOTA methods on the skin lesion segmentation datasets ISIC 2018, ISIC 2017, and ISIC 2016.}
    \adjustbox {max width=\textwidth}
    {
    \begin{tabular}{lccccccccccccccccc}
    \toprule
    \multirow{3}[6]{*}{\textbf{Method}} & \multicolumn{17}{c}{\textbf{Performance (\%)}} \\
\cmidrule{2-18}          & \multicolumn{5}{c}{\textbf{ISIC 2018}} &       & \multicolumn{5}{c}{\textbf{ISIC 2017}} &       & \multicolumn{5}{c}{\textbf{ISIC 2016}} \\
\cmidrule{2-6}\cmidrule{8-12}\cmidrule{14-18}          & $J$ & $D$  & $A_{cc}$   & $S_{n}$   & $S_{p}$    &       & $J$ & $D$  & $A_{cc}$   & $S_{n}$   & $S_{p}$    &       & $J$ & $D$  & $A_{cc}$   & $S_{n}$   & $S_{p}$ \\
\cmidrule{1-6}\cmidrule{8-12}\cmidrule{14-18}    U-Net \cite{ronneberger2015u}  & 80.09 & 86.64 & 92.52 & 85.22 & 92.09 &       & 75.69 & 84.12 & 93.29 & 84.30 & 93.41 &       & 81.38 & 88.24 & 93.31 & 87.28 & 92.88 \\
    UNet++ \cite{zhou2018unet++} & 81.62 & 87.32 & 93.72 & 88.70 & 93.96 &       & 78.58 & 86.35 & 93.73 & 87.13 & 94.41 &       & 82.81 & 89.19 & 93.88 & 88.78 & 93.52 \\
    BCDU-Net \cite{azad2019bi} & 81.10 & 85.10 & 93.70 & 78.50 & 98.20 &       & 79.20 & 78.11 & 91.63 & 76.46 & 97.09 &       & 83.43 & 80.95 & 91.78 & 78.11 & 96.20 \\
    Separable-Unet \cite{TANG2019289} & -     & -     & -     & -     & -     &       & -     & -     & -     & -     & -     &       & 84.27 & 89.95 & 95.67 & 93.14 & 94.68 \\
    CPFNet \cite{9049412} & 79.88 & 87.69 & 94.96 & 89.53 & 96.55 &       & -     & -     & -     & -     & -     &       & 83.81 & 90.23 & 95.09 & 92.11 & 95.91 \\
    DAGAN \cite{LEI2020101716}  & 81.13 & 88.07 & 93.24 & 90.72 & 95.88 &       & 75.94 & 84.25 & 93.26 & 83.63 & 97.25 &       & 84.42 & 90.85 & 95.82 & 92.28 & 95.68 \\
    FAT-Net \cite{WU2022102327} & 82.02 & 89.03 & 95.78 & 91.00 & 96.99 &       & 76.53 & 85.00 & 93.26 & 83.92 & 97.25 &       & 85.30 & 91.59 & 96.04 & 92.59 & 96.02 \\
    AS-Net \cite{HU2022117112}  & 83.09 & 89.55 & 95.68 & 93.06 & 94.69 &       & 80.51 & 88.07 & 94.66 & 89.92 & 95.72 &       & -     & -     & -     & -     & - \\
    SLT-Net \cite{FENG2022105942} & 71.51 & 82.85 & -     & 78.85 & \textbf{99.35} &       & 79.87 & 67.90 & -     & 73.63 & 97.27 &       & -     & -     & -     & -     & - \\
     Ms RED \cite{DAI2022102293} & 83.86 & 90.33 & 96.45 & 91.10 & -     &       & 78.55 & 86.48 & 94.10 & -     & -     &       & 87.03 & 92.66 & 96.42 & -     & - \\
    ARU-GD \cite{maji2022attention} & 84.55 & 89.16 & 94.23 & 91.42 & 96.81 &       & 80.77 & 87.89 & 93.88 & 88.31 & 96.31 &       & 85.12 & 90.83 & 94.38 & 89.86 & 94.65 \\
    EAM-CPFNet\cite{zuo2022efficient}  & 84.58 & 90.81 & 97.10 & -     & -     &       & -     & -     & -     & -     & -     &       & -     & -     & -     & -     & - \\
    ICL-Net \cite{cao2022icl}   & 83.76 & 90.41 & 97.24 & 91.66 & 98.63 &       & -     & -     & -     & -     & -     &       & -     & -     & -     & -     & - \\
    Swin-Unet \cite{cao2023swin} & 82.79 & 88.98 & 96.83 & 90.10 & 97.16 &       & 80.89 & 81.99 & 94.76 & 88.06 & 96.05 &       & 87.60 & 88.94 & 96.00 & 92.27 & 95.79 \\
    \midrule
    \textbf{TBConvL-Net} & \textbf{91.65} & \textbf{95.47} & \textbf{97.6} & \textbf{95.29} & 98.55 &       & \textbf{84.8} & \textbf{90.89} & \textbf{96.07} & \textbf{91.19} & \textbf{97.61} &       & \textbf{89.47} & \textbf{95.45} & \textbf{97.05} & \textbf{94.02} & \textbf{97.68} \\
    \bottomrule
    \end{tabular}%
    }
  \label{tab:ISIC}%
\end{table*}%

\begin{figure*}[h]
	\centering
    \includegraphics[width=\textwidth]{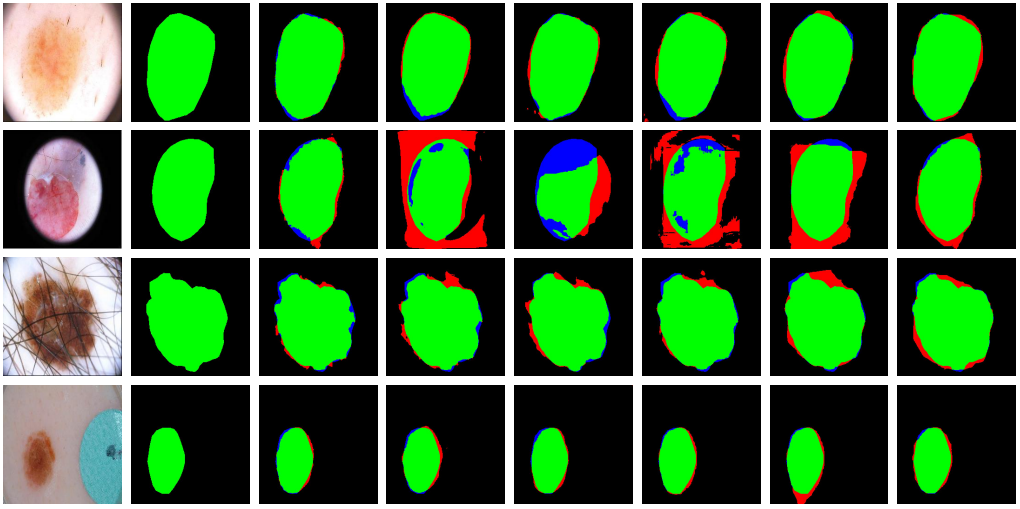} 
	\caption{Example segmentation results of TBConvL-Net on the skin lesions dataset ISIC 2017. From left to right, the columns show the input images, the ground-truth masks, the segmentation results of TBConvL-Net, and the results of ARU-GD \cite{maji2022attention}, UNet++ \cite{zhou2018unet++}, U-Net \cite{ronneberger2015u}, BCDU-Net \cite{azad2019bi}, and Swin-Unet \cite{cao2023swin}, respectively. True-positive pixels are depicted in green, false-positive pixels in red, and false-negative pixels in blue.}
 \label{Fig: Skin Vis}%
\end{figure*}%

In our proposed network, the output of the batch normalisation step, $\beta_{N}^\text{out}$, is fed into a ConvLSTM layer (Fig.~\ref{Fig:CONVLSTM}). This layer comprises a memory cell ($M_{c_t}$), an output gate ($\varnothing_{t}$), an input gate ($i_{t}$) and a forget gate ($f_{t}$). These gates serve as control mechanisms for the ConvLSTM layer, with the input, output, and forget gates specifically controlling the access, updating, and clearing of the memory cells, respectively. The structure and operation of ConvLSTM can be formalised as follows.
\begin{equation}
M_{c_t} = f_{t}\otimes M_{c_{(t-1)}}+i_{t} \tanh (\varpi_{(\Im,M_c)}\circledast \Im _{t} + \varpi_{(h,M_c)}\circledast  \wp _{(t-1)}+\beta_{M_c}),
\label{Eq:mcgate}
\end{equation}\vspace{-1.2\baselineskip}
\begin{equation}
\varnothing_{t}=\varrho(\varpi_{(\Im,\varnothing)}\circledast \Im _{t}+\varpi_{(h,\varnothing)}\circledast \wp _{(t-1)}+\varpi_{(M_c,\varnothing)}\otimes  M_{c_t}+\beta_{\varnothing}),
\label{Eq:outputgate}
\end{equation}
\begin{equation}
i_{t}=\varrho(\varpi_{(\Im,i)}\circledast \Im _{t}+ \varpi_{(h,i)}\circledast  \wp _{(t-1)} +\varpi_{(M_c,i)}\circledast  M_{c_{(t-1)}}+\beta_{i}),
\label{Eq:inputgate}
\end{equation}
\begin{equation}
f_{t}=\varrho(\varpi_{(\Im, f)}\circledast \Im _{t}+ \varpi_{(h,f)}\circledast  \wp _{(t-1)} +\varpi_{(M_c,f)}\circledast  M_{c_{(t-1)}}+\beta_{f}),
\label{Eq:fgate}
\end{equation}
\begin{equation}
\wp _{t}=\varnothing_{g}\otimes \tanh (M_{c_t}),
\label{Eq:hiddentensor}
\end{equation}
where $\otimes$ and $\circledast$ stand for Hadamard and convolutional operations, respectively.
The input and hidden tensors are denoted by $\Im_{t}$ and $\wp_{t}$, respectively. 2D convolution masks of the hidden and input states are denoted by $\varpi_{(\Im, *)}$ and $\varpi_{(h, *)}$. The bias terms of the memory cell, output, input, and forget gates are denoted by $\beta_{M_c}$, $\beta_{\phi}$, $\beta_i$, and $\beta_f$, respectively.

In TBConvL-Net, we use BConvLSTM \cite{song2018pyramid}, which extends traditional ConvLSTM by capturing forward and backward temporal dependencies. This is useful when understanding the past and future context is crucial to interpreting current input features. In a BConvLSTM, input data is processed in two separate paths: a forward and a backward direction, each with its own ConvLSTM layers that process data sequentially. The forward path processes the input data in its natural order, from the first to the last image. The backward path processes the data in reverse, from the last image to the first image. This facilitates the capture of information from both the preceding and subsequent frames with respect to the current input. Studies have shown that considering both forward and backward views improves prediction performance \cite{cui2018deep}. The output of the BConvLSTM is computed as:
\begin{equation}
\Im_\text{out}=\tanh ((\varpi_{}^{\wp\overrightarrow{}}\circledast \wp_{t}^{\overrightarrow{}})+(\varpi_{}^{\wp\overleftarrow{}}\times \wp_{t}^{\overleftarrow{}}) + \beta),
\label{Eq:out}
\end{equation}
where $\wp\overrightarrow{}$ and $\wp\overleftarrow{}$ represent the hidden state tensors for the forward and backward states, respectively, and $\beta$ is the bias component. The hyperbolic tangent function ($\tanh$) is used to non-linearly combine the output of the forward and backward states in BConvLSTM. This ensures effective integration of information from both directions and helps capture complex relationships between the forward and backward dependencies in the input data.


The Swin transformer blocks (Fig.~\ref{Fig: tranformer}) used in our proposed network partition the input into nonoverlapping patches using ViT. Each patch, which encapsulates a $4\times4$ pixel area in our implementation, is treated as a ``token'', its associated features being a combination of the RGB pixel values. These features of the raw value are then projected onto a chosen dimension, denoted by $d$, using a linear embedded layer (LE). Subsequently, a sequence of transformer blocks, equipped with modified self-attention computations, is applied to these patch tokens. This allows the model to learn more complex relationships between input features, which leads to better performance on various tasks. Block 1 consists of transformer blocks and LE, which preserves the token count of ($h/4\times w/4$). As the network progresses, the layers combine to create a hierarchical representation by reducing the token count. The initial patch merging layer (PML) consolidates features from clusters of adjacent $2 \times 2$ patches, after which a linear layer is applied to the 4-$d$-dimensional combined features. This procedure results in a quartering of the token count, which is equivalent to a downsampling of the resolution $2\times$, with the output dimension set to $2\times d$. Subsequently, the transformers are deployed to alter the features while preserving the resolution at ($h/4 \times w/8$). This beginning stage of patch merging and feature conversion is designated as Block 2.

To improve the efficiency of the modelling, self-attention is applied within local windows \cite{liu2021swin}. Given that each window is made up of $N \times N$ patches, the computational complexity of a global multihead self-attention (MSA) module and a shifted window-based MSA (SW-MSA) module for an image of $h \times w$ patches are large.
\begin{equation}
C _\text{MSA}= 4(h\times w) d^{2}+2(h \times w)^{2} d 
\label{Eq:MSA}
\end{equation}
and
\begin{equation}
C_\text{SW-MSA} = 4(h \times w) d^{2} + 2N^{2}(h\times w)^{2} d,
\label{Eq:WMSA}
\end{equation}
respectively, where the former is quadratic in relation to the number of patches and the latter is linear for a fixed $N$. Global self-attention computation is often prohibitively expensive for large $h\times w$, whereas shifted window-based self-attention is scalable. In creating a streamlined version of the Swin Transformers, we substituted the MSA module with a SW-MSA module in each transformer block, retaining the configurations of the remaining layers (see the magnified portion of Fig.~\ref{Fig: tranformer}). This lighter version preserves the essential features of the Swin Transformers while decreasing computational complexity. The overall process of two consecutive Swin Transformer blocks (STBs) is as follows. Let $\tau _\text{in}$ be the input to the first STB and $z_{1}$ be the result of concatenating $\tau _\text{in}$ with the output of the first MSA module after applying a layer norm ($L_{N}$) operation:
\begin{equation}
z_{1}= \tau _\text{in} \copyright L_{N}(\text{MSA}(\tau_\text{in})).
\label{Eq:ST1}
\end{equation}
Next, the input $z_{2}$ to the second STB is calculated as the concatenation of $z_{1}$ and the result of processing $z_1$ by $L_N$ and a multilayer perceptron (MLP):
\begin{equation}
z_{2}= z_{1} \copyright \text{MLP}(L_\text{N}(z_1)).
\label{Eq:ST2}
\end{equation}
In the second STB, $z_{3}$ is calculated by concatenating $z_{2}$ with the output of the SW-MSA module after applying $L_{N}$:
\begin{equation}
z_{3}= z_{2} \copyright L_\text{N}(\text{SW-MSA}(z_{2})).
\label{Eq:ST4}
\end{equation}
Finally, the output $\tau _\text{out}$ of the second STB is calculated as the concatenation of $z_{3}$ and the result of processing $z_3$ by $L_N$ and MLP:
\begin{equation}
\tau _\text{out}= z_3 \copyright \text{MLP}(L_\text{N}(z_3)).
\label{Eq:ST5}
\end{equation}

\begin{figure*}[h]
	\centering
    \includegraphics[width=\textwidth]{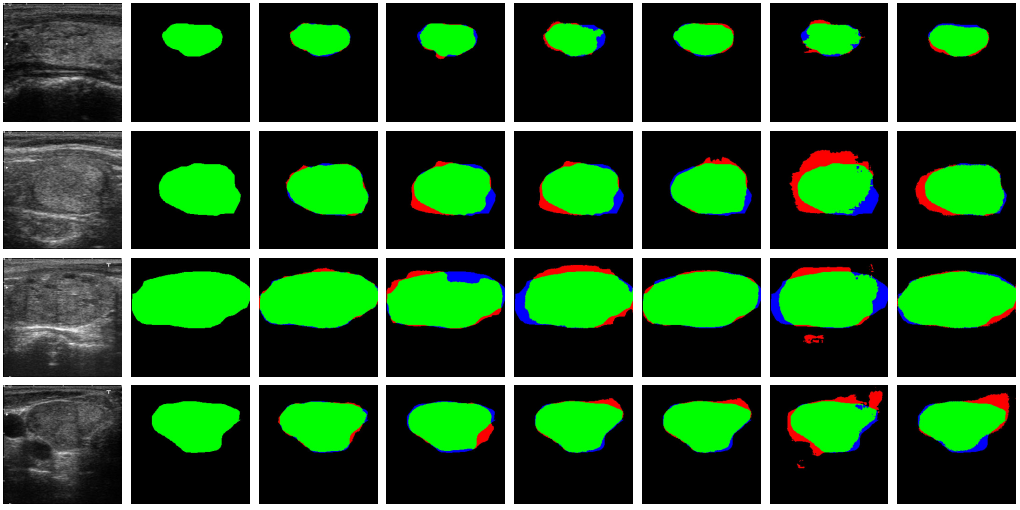} 
	\caption{Example segmentation results of TBConvL-Net on the thyroid nodule dataset DDTI. From left to right, the columns show the input images, the ground-truth masks, the segmentation results of TBConvL-Net, and the results of ARU-GD \cite{maji2022attention}, UNet++ \cite{zhou2018unet++}, U-Net \cite{ronneberger2015u}, BCDU-Net \cite{azad2019bi}, and Swin-Unet \cite{cao2023swin}, respectively. True-positive pixels are depicted in green, false-positive pixels in red, and false-negative pixels in blue.}

	\label{Fig: DDTI Vis}
\end{figure*}
\begin{table}[h]
    \centering
    \caption{Performance comparison of TBConvL-Net with various SOTA methods on the thyroid nodule segmentation dataset DDTI.}
    \adjustbox{max width=\textwidth}{%
    \begin{tabular}{lccccc}
    \toprule
    \multirow{2}[4]{*}{\textbf{Method}} & \multicolumn{5}{c}{\textbf{Performance (\%)}} \\
    \cmidrule{2-6}          & $J$ & $D$  & $A_{cc}$   & $S_{n}$   & $S_{p}$ \\
    \midrule
    U-Net \cite{ronneberger2015u}  & 74.76 & 84.08 & 96.55 & 85.50 & 97.57 \\
    M-Net \cite{mehta2017m} & 79.38 & 86.40 & -     & 75.45 & - \\
    Attention U-Net \cite{oktay2018attention} & 77.37 & 84.91 & -     & 81.70 & - \\
    DeeplabV3+ \cite{chen2017rethinking} & 82.66 & 87.72 & -     & 79.54 & - \\
    UNet++ \cite{zhou2018unet++} & 74.76 & 84.08 & 96.55 & 85.50 & 97.57 \\
    BCDU-Net \cite{azad2019bi} & 57.79 & 69.49 & 93.22 & 78.31 & 94.34 \\
    nnUnet \cite{isensee2021nnu} & 80.76 & 88.59 & -     & 85.23 & - \\
    ARU-GD \cite{maji2022attention} & 77.07 & 83.64 & 97.94 & 83.80 & 98.78 \\
    N-Net \cite{nie2022n} & 88.46 & 92.67 & -     & 91.94 & - \\
    Swin U-Net \cite{cao2023swin} & 75.44 & 84.86 & 96.93 & 86.42 & 97.98 \\
    MShNet \cite{peng2023mshnet} & 73.43 & 75.01 & -     & 82.21 & - \\
    \midrule
    \textbf{TBConvL-Net} & \textbf{88.70} & \textbf{93.56} & \textbf{98.62} & \textbf{94.02} & \textbf{99.09} \\
    \bottomrule
    \end{tabular}%
    }
    \label{tab:DDTI}
\end{table}

\subsection{Loss Function}
\label{sec:Loss}
TBConvL-Net uses ground truth (GT) to supervise the complete segmentation method. The network is trained using a linear combination of Dice loss ($\zeta_{d}$), Jaccard loss ($\zeta_{j}$), and surface boundary loss ($\zeta_b$). One of the main reasons for combining the Dice loss ($\zeta_{d}$) and Jaccard loss ($\zeta_{j}$) is that the former ensures that predictions capture the pathology's overall size and shape, even if it is slightly shifted, while the latter ensures that predictions closely match the shape and location of the pathology. When combining both losses, TBConvL-Net learns to be accurate in terms of both region similarity ($\zeta_{d}$) and placement ($\zeta_{j}$), leading to more precise and robust medical image segmentation.

The Dice loss evaluates the amount of overlap between the segmented image $S$ and the GT image $G$:
\begin{equation}
\zeta_{d}(S,G)=1-\sum_{k=1}^{c}\frac{2w_{k}\sum_{j=1}^{n}S(k,j)\times G(k,j)}{\sum_{j=1}^{n}S(k,j)^{2}+\sum_{j=1}^{n}G(k,j)^{2}}+\xi,
\label{Eq:Diceloss}
\end{equation}
where $w_{k}$ denotes the $k^\text{th}$ class weight, $c$ is the number of classes, $n$ the number of pixels, and $\xi$ is a smoothing constant. The Jaccard loss is calculated as:
\begin{equation}
\zeta_{j}(S,G)=1-\text{IoU}(S,G)-\frac{\left | B- (S\cup G)\right |}{\left | B \right |}+\xi,
\label{Eq:IoUloss}
\end{equation}
where IoU denotes the intersection over union of the segmented image $S$ and the GT image $G$, and $B$ is the bounding box covering $S$ and $G$.

The main purpose of MIS is to accurately identify the edges or boundaries of a lesion. To achieve this, we use a special boundary loss \cite{kervadec2019boundary}. It computes the distance $\text{dist}(\partial S,\partial G)$ between the boundary $\partial S$ of the segmentation mask and the boundary $\partial G$ of the GT mask by integration over the interface where the regions of the two boundaries do not align:
\begin{align}
\text{dist}(\partial S,\partial G) & = \int_{\partial G}\left \| p_{\partial S}(B_{p})- B_{p} \right \|^{2}dB_{p}\label{Eq:B1}\\
& = 2\int_{\Delta S} D_{G}(B_{p})dB_{p}\label{Eq:B2}\\
& = 2\left ( \int_{\Omega }\vartheta _{G}(B_{p})s(B_{p})dB_{p}-\int_{\Omega }\vartheta _{G}(B_{p})g(B_{p})dB_{p} \right )\label{Eq:B3},
\end{align}
where $B_{p}$ is a point on the boundary $\partial G$ and $p_{\partial S}(B_{p})$ is the corresponding point on the boundary $\partial S$, $D_{G}(B_{p})$ is the distance map of point $p$ with respect to the boundary $\partial G$, $\Omega$ denotes the region covered by $S$, $\vartheta _{G}$ is the level-set representation of $\partial G$, calculated as $\vartheta _{G}(p)=-D_{G}(p)$ if $p\in G$ and $\vartheta _{G}=+D_{G}(p)$ otherwise. When $S = S_{\theta}$, the binary variables $s(\cdot)$ in (\ref{Eq:B3}) can be substituted with the softmax probability output of the network, $S_{\theta}(p)$. This leads to the formulation of the boundary loss, which approximates the boundary distance $\text{dist}(\partial S,\partial G)$, subject to a constant that is independent of $\theta$:
\begin{equation}
\zeta_{b}(S,G)=\int_{\Omega }\vartheta _{G}(p)S_{\theta}(p)dp.
\label{Eq: Boundary loss}
\end{equation}

The total loss used to train TBConvL-Net is a linear combination of the Dice, Jaccard, and boundary loss functions:
\begin{equation}
\zeta=\lambda_{d} \zeta_{d}(S,G)+ \lambda_{j} \zeta_{j}(S,G)+ \lambda_{b} \zeta_{b}(S,G),
\label{Eq:Loss}
\end{equation}
where $\lambda_{d}$, $\lambda_{j}$, and $\lambda_{b}$ are the weights (hyperparameters) of the respective loss functions. In our experimentation, we set $\lambda_{d}$ and $\lambda_{j}$ to 1. $\lambda_{b}$ was initialised at 1, then gradually reduced by 0.01 per epoch until it converged at 0.01. This is done to moderate the influence of $\lambda_{b}$ on the boundary constraints, ensuring that its effect remains substantial without excessively dominating the optimisation process.

\section{Experiments and Results}
\label{experimentalResults}

\subsection{Datasets}

The proposed TBConvL-Net model was evaluated on ten challenging benchmark datasets of seven different medical imaging modalities (Table \ref{tab:Datasets}), namely ISIC 2016 \cite{gutman2016skin}, ISIC 2017 \cite{codella2018skin}, and ISIC 2018 \cite{codella2019skin} for the segmentation of skin lesions in optical images, DDTI \cite{DDTI} for the segmentation of thyroid nodules and BUSI \cite{BUSIdataset} for the segmentation of breast cancer in ultrasound images, MoNuSeg for the segmentation of cell nuclei in histopathological whole-slide images, MC \cite{MC_Dataset} for the segmentation of chest X-ray images, IDRiD \cite{IDRiD_Dataset} for the segmentation of the optic disk in fundus images, Fluorescent Neuronal Cells \cite{Fluocell_Dataset} for the segmentation of cells in microscopy images, and TCIA \cite{BrainMRI_Dataset} for the segmentation of brain tumours in magnetic resonance images. All datasets are publicly available and provide GT masks for the evaluation of image segmentation methods. Performance evaluation on the DDTI and BUSI datasets was performed using a five-fold cross-validation method due to the unavailability of a separate test set.

\begin{table}[h]
  \centering
  \caption{Performance comparison of TBConvL-Net model with various SOTA methods on the breast lesion segmentation dataset BUSI.}
    \adjustbox{max width=\textwidth}{%
    \begin{tabular}{lccccc}
    \toprule
    \multirow{2}[4]{*}{\textbf{Method}} & \multicolumn{5}{c}{\textbf{Performance (\%)}} \\
    \cmidrule{2-6} & $J$ & $D$  & $A_{cc}$   & $S_{n}$   & $S{p}$ \\
    \midrule
    U-Net \cite{ronneberger2015u}  & 67.77 & 76.96 & 95.48 & 78.33 & 96.13 \\
    FPN \cite{lin2017feature} & 74.09 & 82.67 & -     & 85.39 & - \\
    DeeplabV3+ \cite{chen2017rethinking} & 73.48 & 82.68 & -     & 83.37 & - \\
    ConvEDNet \cite{lei2018segmentation} & 73.57 & 82.70 & -     & 85.51 & - \\
    UNet++ \cite{zhou2018unet++} & 76.85 & 76.22 & 97.97 & 78.61 & 98.86 \\
    BCDU-Net \cite{azad2019bi} & 74.49 & 66.75 & 94.82 & 86.85 & 95.57 \\
    BGM-Net \cite{wu2021bgm} & 75.97 & 83.97 & -     & 83.45 & - \\
    ARU-GD \cite{maji2022attention} & 77.07 & 83.64 & 97.94 & 83.80 & 98.78 \\
    Swin-Unet \cite{cao2023swin} & 77.16 & 84.45 & 97.55 & 84.81 & 98.34 \\
    \midrule
    \textbf{TBConvL-Net} & \textbf{91.97} & \textbf{95.72} & \textbf{99.50} & \textbf{95.85} & \textbf{99.69} \\
    \bottomrule
    \end{tabular}%
  }
  \label{tab:BUSI}
\end{table}

\begin{figure*}[h]
	\centering
    \includegraphics[width=\textwidth]{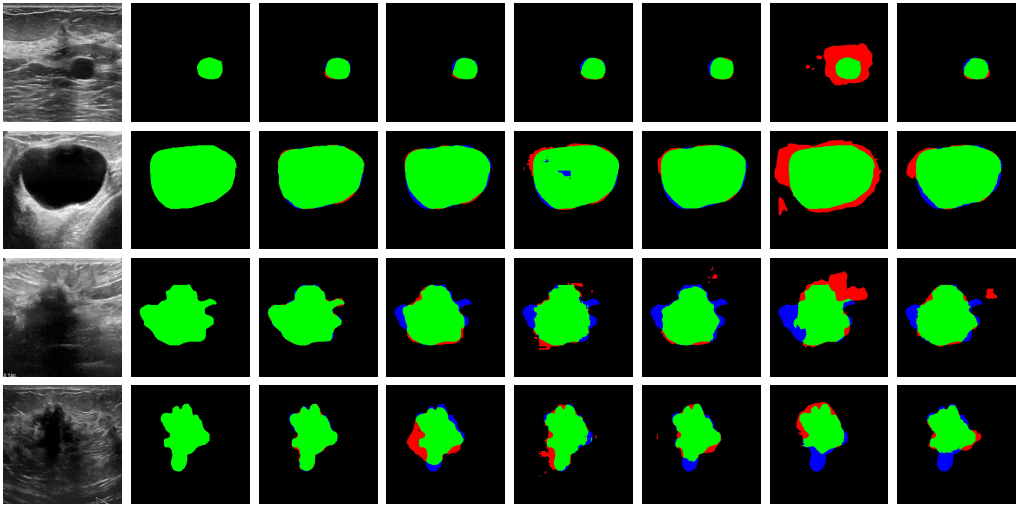}
	\caption{Example segmentation results of TBConvL-Net on the breast lesion dataset BUSI. From left to right, the columns show the input images, the ground-truth masks, the segmentation results of TBConvL-Net, and the results of ARU-GD \cite{maji2022attention}, UNet++ \cite{zhou2018unet++}, U-Net \cite{ronneberger2015u}, BCDU-Net \cite{azad2019bi}, and Swin-Unet \cite{cao2023swin}. True-positive pixels are depicted in green, false-positive pixels in red, and false-negative pixels in blue.}
    \label{Fig: BUSI Vis}
\end{figure*}
    
\subsection{Evaluation Criteria}

The segmentation performance of TBConvL-Net was evaluated and compared with SOTA methods using several metrics, including the Jaccard index ($J$, equal to IoU), Dice similarity coefficient ($D$), accuracy (A$_{cc}$), sensitivity (S$_n$), and specificity (S$_p$). These metrics were calculated as per their definitions:

\begin{equation}
J =\frac{T_{P}}{T_{P}+F_{P}+F_{N}},
\label{eq:Jindex}
\end{equation}

\begin{equation}
D =\frac{2 \times T_{P}}{2\times T_{P}+F_{P}+F_{N}},
\label{eq:Dice}
\end{equation}

\begin{equation}
A_{cc}=\frac{T_{P}+T_{N}}{T_{P}+T_{N}+F_{P}+F_{N}},
\label{eq:acc}
\end{equation}

\begin{equation}
S_{n}=\frac{T_{P}}{T_{P}+F_{N}},
\label{eq:sn}
\end{equation}

\begin{equation}
S_{p}=\frac{T_{N}}{T_{N}+F_{P}},
\label{eq:sp}
\end{equation}

\noindent where $T_{P}$, $T_{N}$, $F_{P}$, and $F_{N}$ denote the number of true positives, true negatives, false positives, and false negatives, respectively.

\subsection{Training Details}

For model training, the images (Table \ref{tab:Datasets}) were augmented using contrast adjustments (with factors of [$\times 0.9, \times 1.1$]) and flipping operations (both in horizontal and vertical directions), which increased the size of the datasets by a factor of 5. Segmentation models were trained through various mixtures of loss functions and training methodologies. The Adam optimiser was used with a maximum of 60 iterations and an initial learning rate of 0.001. In the absence of performance improvement on the validation set after five epochs, the learning rate was reduced by a quarter. To stop overfitting, an early stop strategy was implemented. The models were implemented through Keras using TensorFlow as the back-end and trained on an NVIDIA K80 GPU.

\subsection{Ablation Experiments}

To evaluate the impact of the main components, loss functions, and training strategies used in TBConvL-Net, three ablation experiments were carried out.

The first ablation experiment was conducted using the ISIC 2017 dataset, as it is one of the more challenging datasets. The experiment began with a simple bidirectional ConvLSTM U-Net (BCDU-Net \cite{azad2019bi}) as the baseline model (BM), and then traditional convolutional layers were replaced with depth-wise separable convolutions, resulting in substantial reductions in computational costs. The filters were then optimised. The Swin Transformer was then used at various locations within the network. We found that performance improved substantially when the Swin Transformer was used between the skip connections and the deeply separable convolutional layer, densely connected in depth, of the network (Table~\ref{tab:ablation 1}). All results of this first ablation were computed using only the Dice loss.

The second ablation experiment was conducted to understand the influence of different loss functions and was performed on the DDTI dataset. A variety of loss functions, such as dice loss, Jaccard loss, and boundary loss, were examined individually and in various combinations. Given that the DDTI dataset comprises images with irregular shapes and boundaries, we discovered that the linear combination of Dice loss, Jaccard loss, and boundary loss yielded the best performance, both quantitatively (Table~\ref{tab:ablation 2}) and qualitatively (Fig.~\ref{Fig: DDTI Loss Ablation}). Thus, we used this combined loss in all subsequent experiments.

Finally, in the third ablation experiment, we investigated the potential of transfer learning to further boost the performance of the proposed TBConvL-Net. The rationale of this experiment was to capitalise on the principle that transferring domain knowledge from other modalities, meaning incorporating preexisting feature representations learnt from these other sources, may be beneficial to the segmentation process and yield better results for any given dataset. For each dataset (Table~\ref{tab:Datasets}) we experimented with transfer learning from other selected datasets (Table~\ref{tab:trainig_strategy}) to facilitate feature learning. For some datasets, this involved sequential learning, as in the case of the ISIC 2016 dataset, where weights were first learnt from the ISIC 2017 dataset, and then the model was further trained on the ISIC 2016 dataset. For other datasets, specifically MC and IDRiD, we found that our model already attained state-of-the-art performance without employing transfer learning. We observed that transfer learning generally improves segmentation performance or otherwise does not decrease performance (Table~\ref{tab:Ablation 3}). Thus, for comparisons with other state-of-the-art methods presented next, we used this transfer learning strategy and the combined loss function.
\begin{table}[h]
    \centering
    \caption{Performance comparison of TBConvL-Net model with various SOTA methods on the cell nuclei segmentation dataset MoNuSeg and the fluorescent neuronal cell segmentation dataset Fluorescent Neuronal Cells.}
    \adjustbox {max width=\textwidth}{%
    \begin{tabular}{lccccccccccc}
    \toprule
    \multirow{3}[6]{*}{\textbf{Method}} & \multicolumn{11}{c}{\textbf{Performance (\%)}} \\
    \cmidrule{2-12}          & \multicolumn{5}{c}{\textbf{MoNuSeg}}  &       & \multicolumn{5}{c}{\textbf{Fluoscent Neuronal Cells}} \\
    \cmidrule{2-6}\cmidrule{8-12}          & \textbf{$J$} & \textbf{$D$} & \textbf{$A_{cc}$} & \textbf{$S_{n}$} & \textbf{$S_{p}$} &       & \textbf{$J$} & \textbf{$D$} & \textbf{$A_{cc}$} & \textbf{$S{n}$} & \textbf{$S_{p}$} \\
    \midrule
    U-Net \cite{ronneberger2015u}  & 62.16 & 75.48 & 90.24 & 81.17 & 92.02 &       & 74.48 & 84.63 & 99.53 & 81.57 & 99.82 \\
    UNet++ \cite{zhou2018unet++} & 62.05 & 75.30 & 89.79 & 81.32 & 91.49 &       & 70.99 & 81.97 & 99.47 & 78.75 & 99.81 \\
    BCDUnet \cite{azad2019bi} & 66.61 & 79.82 & 92.05 & 82.48 & 94.12 &       & 74.80 & 84.79 & 99.53 & 82.34 & 99.83 \\
    ARU-GD \cite{maji2022attention} & 60.76 & 73.89 & 90.97 & 75.54 & 93.70 &       & 66.27 & 78.74 & 99.35 & 75.24 & 99.77 \\
    c-ResUnet \cite{morelli2021automating} & 66.61 & 79.82 & 92.05 & 82.48 & 94.12 &       & 82.03 & 89.97 & 99.69 & 82.46 & 99.99 \\
    \midrule
    \textbf{TBConvL-Net} & \textbf{76.07} & \textbf{85.16} & \textbf{93.62} & \textbf{88.04} & \textbf{95.53} &       & \textbf{92.84} & \textbf{96.23} & \textbf{99.90} & \textbf{97.01} & \textbf{99.94} \\
    \bottomrule
    \end{tabular}%
    }
  \label{tab:MoNuSeg_Fluocell}
\end{table}

\begin{figure*}[h]
	\centering
    \includegraphics[width=0.9\textwidth]{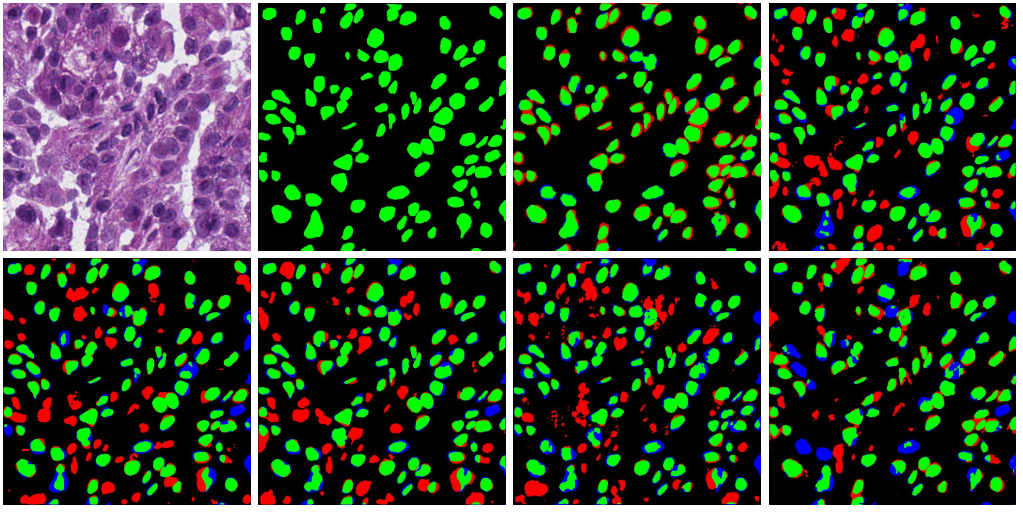} 
    \caption{Example segmentation results of TBConvL-Net on the MoNuSeg dataset. From top-left to bottom-right, the panels show the input image, the corresponding ground-truth mask, the segmentation result of TBConvL-Net, and the results of U-Net \cite{ronneberger2015u}, UNet++ \cite{zhou2018unet++}, BCDU-Net \cite{azad2019bi}, ARU-GD \cite{maji2022attention}, and c-ResUnet \cite{morelli2021automating}. True-positive pixels are depicted in green, false-positive pixels in red, and false-negative pixels in blue.}
    \label{Fig: MoNuSeg}
\end{figure*}

\begin{figure*}[h]
	\centering
    \includegraphics[width=0.9\textwidth]{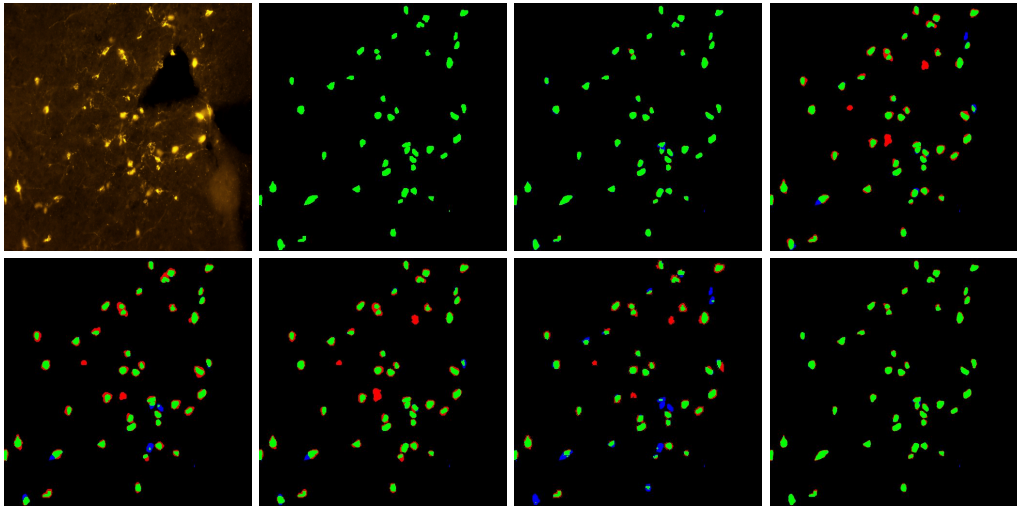}
	\caption{Example segmentation results of TBConvL-Net on the Fluorescent Neuronal Cells dataset. From top-left to bottom-right, the panels show the input image, the corresponding ground-truth mask, the segmentation result of TBConvL-Net, and the results of U-Net \cite{ronneberger2015u}, UNet++ \cite{zhou2018unet++}, BCDU-Net \cite{azad2019bi}, ARU-GD \cite{maji2022attention}, and c-ResUnet \cite{morelli2021automating}. True-positive pixels are depicted in green, false-positive pixels in red, and false-negative pixels in blue.}
	\label{Fig: Floucell}
\end{figure*}

\subsection{Comparisons With State-of-the-Art Methods}
\label{sec:Results}

\begin{table}[h]
    \centering
    \caption{Performance comparison of TBConvL-Net model with various SOTA methods on the optic disc segmentation dataset IDRiD and the chest X-ray segmentation dataset MC.}
    \adjustbox {max width=\textwidth}{%
    \begin{tabular}{lccccccccccc}
    \toprule
    \multirow{3}[6]{*}{\textbf{Method}} & \multicolumn{11}{c}{\textbf{Performance (\%)}} \\
    \cmidrule{2-12}          & \multicolumn{5}{c}{\textbf{IDRiD}} &       & \multicolumn{5}{c}{\textbf{MC}} \\
    \cmidrule{2-6}\cmidrule{8-12}          & \textbf{$J$} & \textbf{$D$} & \textbf{$A_{cc}$} & \textbf{$S_{n}$} & \textbf{$S_{p}$} &       & \textbf{$J$} & \textbf{$D$} & \textbf{$A_{cc}$} & \textbf{$S_{n}$} & \textbf{$S_{p}$} \\
    \midrule
    U-Net \cite{ronneberger2015u}  & 90.22 & 94.65 & 99.81 & 94.07 & 99.93 &       & 96.47 & 98.20 & 99.14 & 97.91 & 99.51 \\
    UNet++ \cite{zhou2018unet++} & 87.87 & 92.87 & 99.71 & 94.62 & 99.80 &       & 95.64 & 97.77 & 98.94 & 97.56 & 99.34 \\
    BCDUnet \cite{azad2019bi} & 88.74 & 87.02 & 99.35 & 79.84 & 99.88 &       & 96.39 & 98.16 & 99.11 & 97.82 & 99.50 \\
    ARU-GD \cite{maji2022attention} & 91.59 & 95.57 & 99.85 & 95.30 & 99.93 &       & 96.14 & 98.03 & 99.00 & 97.98 & 99.32 \\
    \midrule
    \textbf{TBConvL-Net} & \textbf{95.65} & \textbf{96.73} & \textbf{99.94} & \textbf{97.68} & \textbf{99.97} &       & \textbf{97.88} & \textbf{98.97} & \textbf{99.50} & \textbf{98.40} & \textbf{99.05} \\
    \bottomrule
    \end{tabular}%
    }
  \label{tab:MC_OD}
\end{table}

\begin{figure*}[h]
	\centering
    \includegraphics[width=\textwidth]{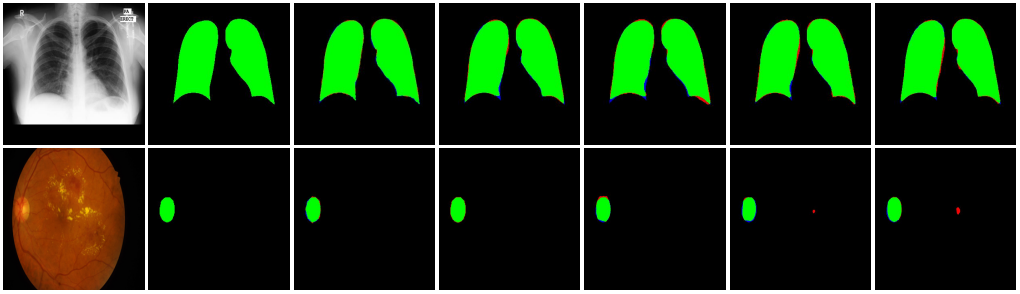}
    \caption{Example segmentation results of TBConvL-Net on the MC dataset (top row) and IDRiD dataset (bottom row). From left to right, the columns show the input images, the ground-truth masks, the segmentation results of TBConvL-Net, and the results of ARU-GD \cite{maji2022attention}, UNet++ \cite{zhou2018unet++}, U-Net \cite{ronneberger2015u}, BCDU-Net \cite{azad2019bi}, and ARU-GD \cite{maji2022attention}. True-positive pixels are depicted in green, false-positive pixels in red, and false-negative pixels in blue.}
	\label{Fig: MC_OD}
\end{figure*}

\begin{table}[h]
    \centering
    \caption{Performance comparison of TBConvL-Net model with various SOTA methods on brain tumour segmentation using the TCIA dataset.}
    \adjustbox {max width=\textwidth}{%
    \begin{tabular}{lccccc}
    \toprule
    \multirow{2}[4]{*}{\textbf{Method}} & \multicolumn{5}{c}{\textbf{Performance (\%)}} \\
    \cmidrule{2-6}          & $J$ & $D$  & $A_{cc}$   & $S_{n}$   & $S_{p}$ \\
    \midrule
    U-Net \cite{ronneberger2015u}  & 86.15 & 90.28 & 98.67 & 89.26 & 99.27 \\
    UNet++ \cite{zhou2018unet++} & 78.44 & 83.42 & 98.22 & 85.69 & 98.88 \\
    BCDU-Net \cite{azad2019bi} & 84.18 & 87.97 & 98.45 & 88.16 & 99.10 \\
    ARU-GD \cite{maji2022attention} & 81.68 & 86.73 & 98.67 & 85.81 & 99.35 \\
    Swin-Unet \cite{cao2023swin} & 83.46 & 87.86 & 98.81 & 91.64 & 99.23 \\
    \midrule
    \textbf{Proposed TBConvL-Net} & \textbf{92.93} & \textbf{95.47} & \textbf{99.34} & \textbf{95.63} & \textbf{99.79} \\
    \bottomrule
    \end{tabular}%
    }
  \label{tab:Brain_MRI}
\end{table}

\begin{figure*}[h]
	\centering
    \includegraphics[width=\textwidth]{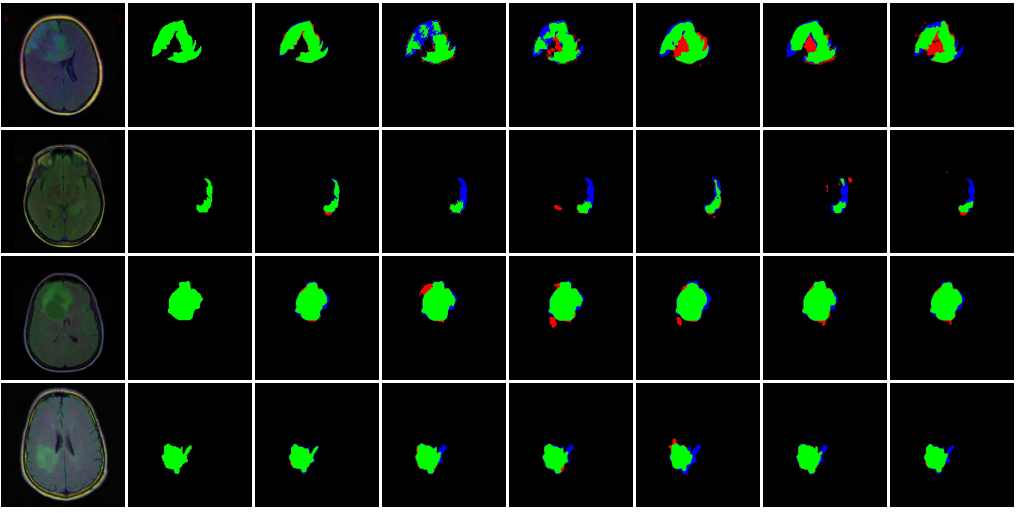} 
    \caption{Example segmentation results of TBConvL-Net on the brain tumour segmentation dataset TCIA. From left to right, the columns show the input images, the corresponding ground-truth masks, the segmentation results of TBConvL-Net, and the results of U-Net \cite{ronneberger2015u}, UNet++ \cite{zhou2018unet++}, ARU-GD \cite{maji2022attention}, BCDU-Net \cite{azad2019bi}, and Swin Unet \cite{cao2023swin}. True-positive pixels are depicted in green, false-positive pixels in red, and false-negative pixels in blue.}
    \label{Fig: Brain_MRI}
\end{figure*}

For each of the datasets (Table~\ref{tab:Datasets}) we compared TBConvL-Net with various SOTA methods. Given the large number of methods and the unavailability of many, it was not feasible to reimplement, retrain, rerun, and/or reevaluate them. Instead, we copied the performance scores reported by the original developers in their papers, as cited throughout this section. This also means that the lists of SOTA methods may be different for each dataset, as in the literature not all methods we compared with were evaluated on all datasets. If scores were not reported for certain metrics, we indicate this with a dash (-) in our tables.

Comparison of TBConvL-Net for skin lesion segmentation in the ISIC 2016, 2017, and 2018 datasets (Table~\ref{tab:ISIC}) shows that our method performed better in terms of virtually all metrics. For example, compared to the SOTA methods listed, TBConvL--Net scored 1. 87\% ----8. 09\%, 3. 91\% ----9. 11\% and 7. 07\% ----20. 14\% better in terms of the Jaccard index in ISIC 2016, 2017, and 2018, respectively. Furthermore, we observed that TBConvL-Net shows better performance in images of skin lesions with various challenges, such as irregular shapes, varying sizes, and the presence of hair, artefacts, and multiple lesions (Fig.~\ref{Fig: Skin Vis}).

Next, a comparison of TBConvL-Net for the segmentation of thyroid nodules in the DDTI dataset (Table~\ref{tab:DDTI}) and the segmentation of breast cancer lesion in the BUSI dataset (Table~\ref{tab:BUSI}) shows that our method performed better in terms of all metrics. For example, compared to the listed SOTA methods, TBConvL-Net scored 0.24\%--30.91\%  and 14.81\%--24.2\% better in terms of the Jaccard index on the DDTI and BUSI datasets, respectively. Furthermore, we observed that TBConvL-Net shows better performance on thyroid nodule image (Fig.~\ref{Fig: DDTI Vis}) and breast lesions (Fig.~\ref{Fig: BUSI Vis}) with various challenges, such as irregular shapes, varying sizes, and the presence of hair, artefacts, and multiple lesions.

Similarly, the comparison of TBConvL-Net for cell nuclei segmentation on the MoNu\-Seg dataset and fluorescent neuronal cell segmentation on the Fluorescent Neuronal Cells dataset (Table~\ref{tab:MoNuSeg_Fluocell}) shows that our method performed superiorly in terms of all metrics. For example, compared to the listed SOTA methods, TBConvL-Net scored 9.46\%--15.31\%  and 10.81\%--26.57\% better in terms of the Jaccard index on the two datasets, respectively. Visual results for some example cell nuclei (Fig.~\ref{Fig: MoNuSeg}) and neuronal cells (Fig.~\ref{Fig: Floucell}) confirm the quantitative results and show that the segmentations closely resemble the GT data, even for images with varying object sizes, irregular shapes, and low contrast.

Furthermore, the comparison of TBConvL-Net for optical disc segmentation in the IDRiD data set and chest X-ray image segmentation in the MC data set (Table~\ref{tab:MC_OD}) shows that our method performed superiorly in terms of all metrics for these tasks also. For example, compared to the listed SOTA methods, TBConvL-Net scored 4.06\%--7.78\%  and 1.41\%--2.24\% better in terms of the Jaccard index on the IDRiD and MC datasets, respectively. This is confirmed by visual examination (Fig.~\ref{Fig: MC_OD}),  which shows that the output of TBConvL-Net closely resembles the GT data, even for images with varying sizes and low contrast.

Finally, the comparison of TBConvL-Net for brain tumour segmentation in the TCIA data set (Table~\ref{tab:Brain_MRI}) again shows that our method performed superiorly in terms of all metrics. For example, compared to the listed SOTA methods, TBConvL-Net scored 6.78\%--14.49\% better in terms of the Jaccard index. Furthermore, we observed that TBConvL-Net shows better performance on images with various challenges, such as irregular shapes, varying sizes, and the presence of hairs, artifacts, and multiple lesions (Fig.~\ref{Fig: Brain_MRI}).

\begin{table}[h]
  \centering
  \caption{Comparison of TBConvL-Net with other SOTA methods in terms of their numbers of parameters, floating-point operations per second (FLOPS), and inference times.}
  \adjustbox {max width=\textwidth}{%
    \begin{tabular}{lccc}
    \toprule
    \multirow{2}[2]{*}{\textbf{Method}} & \textbf{Parameters} & \textbf{FLOPs} & \textbf{Inference Time} \\
    & \textbf{(M)} $\downarrow$ & \textbf{(G) $\downarrow$} & \textbf{(msec) $\downarrow$} \\
    \midrule
    U-Net \cite{ronneberger2015u} & 23.6 & 33.4 & 28.9 \\
    ARU-GD \cite{maji2022attention} & 23.7 & 33.9 & 29.5 \\
    DeeplabV3+ \cite{chen2017rethinking} & 26.2 & 33.9 & 29.6 \\
    UNet++ \cite{zhou2018unet++} & 24.4 & 35.6 & 31.3 \\
    BCDU-Net \cite{azad2019bi}  & 20.7 & 112.0 & 28.1 \\
    Swin UNet \cite{cao2023swin}  & 27.3 & 37.0 & 34.8 \\
    \midrule
    \textbf{TBConvL-Net} & \textbf{9.6} & \textbf{15.5} & \textbf{19.1} \\
    \bottomrule
    \end{tabular}%
    }
  \label{tab: complexity}
\end{table}
\subsection{Models Complexity Analysis}
We also compared the complexity of our proposed TBConvL-Net with other methods in terms of the number of parameters, floating-point operations per second (FLOPs), and inference time (Table \ref{tab: complexity}). TBConvL-Net has 9.6 million (M) parameters, 15.5 billion (G) FLOPs, and an inference time of 19.1 milliseconds (ms). This outperforms all other methods used for visual performance comparisons in all three aspects. Swin Unet \cite{cao2023swin} adopts global self-attention with a transformer structure, leading to high computational costs of 27.3M parameters, 37.0G FLOPs, and an inference time of 34.8 ms, which is 2.84, 2.39, and 1.82 times greater than the proposed TBConvL-Net. Even with its reduced complexity, TBConvL-Net achieves superior segmentation performance compared to Swin Unet \cite{cao2023swin}. The results suggest that our proposed model achieves the best balance between model complexity and segmentation performance.

\section{Conclusions}
\label{sec:Conclusions}
This article introduces a new hybrid deep neural network architecture called TBConvL-Net for MIS tasks. It effectively combines the advantages of CNNs and vision transformers, overcoming the limitations of each technique. The proposed encoder-decoder architecture features depth-wise separable and densely connected convolutions for robust and unique feature learning, network optimisation, and improved generalisation. Additionally, the Bidirectional ConvLSTM and Swin Transformer modules are integrated into skip connections to refine the feature extraction process. The TBConvL-Net model was compared with previous CNNs, transformer-based models, and hybrid approaches. The findings indicate that TBConvL-Net surpasses these models in several MIS tasks by capturing multiscale, long-range dependencies, and local spatial information. Moreover, the proposed model strikes a good balance between complexity and segmentation performance. TBConvL-Net has shown promising results in the domain of MIS, and future experiments could potentially further broaden its range of applications to other areas of medical imaging.

\end{document}


\begin{frontmatter}

\title{TBConvL-Net: A Hybrid Deep Learning Architecture for Robust Medical Image Segmentation}

\author{Shahzaib Iqbal}
\author{Tariq M. Khan}
\author{Syed S. Naqvi}
\author{Asim Naveed}
\author{Erik Meijering}

\end{frontmatter}

\begin{center}
\vspace{-1.75\baselineskip}
\large{Supplementary Material}
\vspace{\baselineskip}
\end{center}

\section{Brief Description of Comparison Methods}

In the presented experiments, we compare TBConvL-Net with several variants of U-Net \cite{ronneberger2015u}, including UNet++ \cite{zhou2018unet++} and BCDU-Net \cite{azad2019bi}, as they are considered benchmark methods for biomedical image segmentation. Specifically, UNet++ extends the pioneer U-Net method by incorporating dense skip connections, deep supervision and improved feature fusion, whereas BCDU-Net extends the U-Net architecture by modeling feature fusion in a nonlinear fashion by incorporating bidirectional information flow and richer feature representations via densely connected convolutions and feature reuse.

To represent the state-of-the-art (SOTA) in biomedical image segmentation, nnUnet \cite{isensee2021nnu} and Swin-UNet \cite{cao2023swin} are also included. nnUnet features task-aware self-configuration, robustness and generalization to a wide variety of tasks, through an automated pipeline. Contrarily, Swin-UNet reforms the U-Net architecture by incorporating a transformer architecture for improved feature extraction and enhanced generalization to diverse medical image segmentation tasks. It replaces fully convolutional and hybrid approaches (combining CNNs and transformers) by a pure transformer-based approach for medical image segmentation.

Other representative SOTA methods included in our experiments are FAT-Net \cite{WU2022102327}, DAGAN \cite{LEI2020101716}, and Ms RED \cite{DAI2022102293}, which were chosen for skin lesion segmentation due to their superior performance and extensive validation in the literature. Furthermore, BGM-Net \cite{wu2021bgm} was chosen for its use of boundary-guided modules (BGMs) to improve segmentation accuracy, especially around object borders. Similarly, ARU-GD \cite{maji2022attention} uses residual connections and attention processes to enhance feature extraction and integration. By utilizing global descriptors to obtain contextual information, ARU-GD successfully handles challenging background problems. c-ResUnet \cite{morelli2021automating} improves feature learning and segmentation accuracy by fusing the U-Net design with residual connections. This network performs better at segmenting complex structures and makes use of residual blocks to enhance gradient flow, which helps training deeper networks. Finally, TSCA-Net \cite{fu2024tsca} combines transformer structures with a context-aware module for improved feature representation to collect global contextual information efficiently. By fusing transformers with conventional convolutional networks, the novel design of TSCA-Net exhibits enhanced segmentation performance, underscoring its potential for useful applications in image processing and medical imaging.

\begin{table}[!t]
  \centering\small
  \caption{Comparison of the p-values from the paired t-test performed on the Jaccard scores of TBConvL-Net with four other available methods on all considered datasets.}
  \adjustbox{max width=\textwidth}{%
    \begin{tabular}{lcccc}
    \toprule
    \multirow{2}[4]{*}{\textbf{Dataset}} & \multicolumn{4}{c}{\textbf{Method}} \\
    \cmidrule{2-5}
    & U-Net \cite{ronneberger2015u}  & UNet++ \cite{zhou2018unet++} & ARU-GD \cite{maji2022attention} & Swin-UNet \cite{cao2023swin} \\
    \midrule
    ISIC 2016 & 0.0001 & 0.0268 & 0.0259 & 0.0119 \\
    ISIC 2017 & 0.0105 & 0.0054 & 0.0180 & 0.0376 \\
    ISIC 2018 & 0.0279 & 0.0032 & 0.0071 & 0.0020 \\
    DDTI      & 0.0067 & 0.0123 & 0.0353 & 0.0366 \\
    BUSI      & 0.0084 & 0.0385 & 0.0353 & 0.0387 \\
    MoNuSeg   & 0.0018 & 0.0020 & 0.0066 & 0.0123 \\
    MC        & 0.0210 & 0.0033 & 0.0032 & 0.0245 \\
    IDRiD     & 0.0399 & 0.0232 & 0.0251 & 0.0177 \\
    FluoCells & 0.0332 & 0.0207 & 0.0379 & 0.0020 \\
    TCIA      & 0.0053 & 0.0182 & 0.0396 & 0.0020 \\
    \bottomrule
    \end{tabular}}
  \label{tab:t_test}
\end{table}

\section{Statistical Significance of TBConvL-Net Results}

To statistically assess the results of TBConvL-Net compared to several SOTA methods, paired t-tests were conducted on the Jaccard scores, and p-values were recorded. This analysis was done only for a subset of methods, namely U-Net, UNet++, ARU-GD, and Swin-UNet, as it required reproducing the results of these methods. The results (Table \ref{tab:t_test}) reveal that when using the typical threshold of p$<$0.05, the improvements in the segmentation results of our method are consistently statistically significant across all datasets.

\section{Evaluation of TBConvL-Net on Multiclass Segmentation}

To explore the potential of TBConvL-Net for multiclass segmentation, we also performed a preliminary experiment on the ACDC\footnote{\tiny\url{https://www.creatis.insa-lyon.fr/Challenge/acdc/databases.html}} dataset, using the Hausdorff distance (HD) and the Dice score (D) as the main metrics, like in the ACDC study. Here again we used U-Net, UNet++, ARU-GD, and Swin-UNet as the comparison methods. The results (Table \ref{tab:ACDC}) suggest that our model achieves superior results also for this task, providing further evidence for its strong generalization capabilities and robustness (see Fig.~\ref{fig:ACDC} for visual examples).

\begin{table}[!t]
  \centering
  \caption{Quantitative comparison of the proposed TBConvL-Net with other methods on the ACDC dataset. Values are presented in mean ± std notation. Best values are indicated in bold. Abbreviations: HD = Hausdorff distance (lower is better), D = Dice score (higher is better), RV = right ventricle, Myo = myocardium, LV = left ventricle.}
  \adjustbox {max width=\textwidth}{%
   \begin{tabular}{lccccccccccc}
    \toprule
    \multirow{3}[6]{*}{\textbf{Method}} & \multicolumn{11}{c}{\textbf{Performance (\%)}} \\
    \cmidrule{2-12} & \multicolumn{2}{c}{\textbf{Average}} && \multicolumn{2}{c}{\textbf{RV}} && \multicolumn{2}{c}{\textbf{Myo}} && \multicolumn{2}{c}{\textbf{LV}} \\
    \cmidrule{2-3}\cmidrule{5-6}\cmidrule{8-9}\cmidrule{11-12}
    & \textbf{HD} & \textbf{D} && \textbf{HD} & \textbf{D} && \textbf{HD} & \textbf{D} && \textbf{HD} & \textbf{D} \\
    \midrule
    U-Net \cite{ronneberger2015u}   & 10.76 ± 4.76 & 79.73 ± 7.48 && 8.60 ± 1.21 & 80.48 ± 8.23 && 12.15 ± 6.77 & 77.70 ± 1.99 && 11.54 ± 6.31 & 79.01 ± 12.22 \\
    UNet++ \cite{zhou2018unet++}    &  7.21 ± 1.82 & 84.45 ± 1.98 && 6.82 ± 2.04 & 89.80 ± 1.63 &&  8.02 ± 2.24 & 82.53 ± 2.98 &&  6.79 ± 1.17 & 84.00 ±  1.33 \\
    ARU-GD \cite{maji2022attention} &  7.42 ± 1.71 & 87.55 ± 4.78 && 7.48 ± 1.09 & 92.74 ± 5.10 &&  8.50 ± 1.70 & 87.09 ± 3.24 &&  6.30 ± 2.36 & 82.82 ±  6.00 \\
    Swin-UNet \cite{cao2023swin}    &  6.05 ± 2.26 & 89.42 ± 3.72 && 6.74 ± 1.09 & 87.90 ± 1.70 &&  5.88 ± 3.44 & 78.87 ± 1.93 &&  5.54 ± 2.25 & 80.47 ±  2.52 \\
    \midrule
    \textbf{Proposed TBConvL-Net} & \textbf{1.98 ± 0.69} & \textbf{90.33 ± 1.08} && \textbf{1.87 ± 0.24} & \textbf{92.78 ± 0.51} && \textbf{2.80 ± 1.53} & \textbf{90.22 ± 1.55} && \textbf{1.26 ± 0.31} & \textbf{88.00 ± 1.20} \\
    \bottomrule
    \end{tabular}}
  \label{tab:ACDC}
\end{table}

\begin{figure}[!t]
    \centering
    \includegraphics[width=\textwidth]{Visual Results/ACDC/ACDC_Visuals.pdf}
    \caption{Example segmentation results on the ACDC dataset. From left to right: (a) input image, (b) the corresponding ground-truth mask, and the segmentation results of (c) TBConvL-Net, (d) Swin-UNet \cite{cao2023swin}, (e) ARU-GD \cite{maji2022attention}, (f) UNet++ \cite{zhou2018unet++}, and (g) U-Net \cite{ronneberger2015u}. Abbreviations: RV = right ventricle, Myo = myocardium, LV = left ventricle.}
    \label{fig:ACDC}
\end{figure}

\footnotesize
\bibliographystyle{elsarticle-num} 
\bibliography{egbib}